\documentclass[aps,prd,10pt,twocolumn,floatfix,nofootinbib,amsmath,amssymb,altaffilletter,preprintnumbers,longbibliography]{revtex4-2}
\pdfoutput=1

\usepackage{graphicx}
\usepackage{dcolumn}
\usepackage{bm}
\usepackage{braket}
\usepackage{dsfont, fontawesome}
\usepackage[colorlinks=true, pdfstartview=FitV,  breaklinks=true]{hyperref}
\usepackage{color}
\usepackage[dvipsnames,table]{xcolor}
\hypersetup{urlcolor=BlueViolet, citecolor=Plum, linkcolor=PineGreen}
\usepackage{cleveref}
\usepackage{comment}
\usepackage{booktabs}
\usepackage{siunitx}
\usepackage{soul}
\usepackage{bm}
\usepackage{esint}
\usepackage{lettrine}
\usepackage{Zallman}
\usepackage{svg}

\makeatletter
\renewcommand{\@seccntformat}[1]{%
  \ifcsname prefix@#1\endcsname
    \csname prefix@#1\endcsname
  \else
    \csname the#1\endcsname\quad
  \fi}

\makeatother

\newcommand{\KCL}{\affiliation{Theoretical Particle Physics and Cosmology Group, Department of Physics, King's College London, UK}}

\begin{document}

\preprint{ \includegraphics[width=0.4\textwidth]{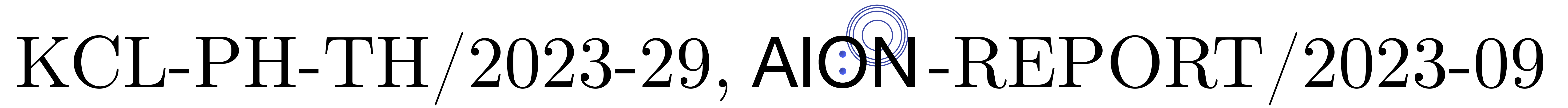}  }

\title{\texorpdfstring{From RATs to riches: mitigating anthropogenic and synanthropic\\ noise in atom interferometer searches for ultra-light dark matter}{}}
\author{John Carlton}
\email{john.carlton@kcl.ac.uk}
\author{Christopher McCabe}
\KCL


\begin{abstract}
 Atom interferometers offer promising new avenues for detecting ultra-light dark matter (ULDM).
The exceptional sensitivity of atom interferometers to fluctuations in the local gravitational potential exposes 
them to sources of noise from human (anthropogenic) and animal (synanthropic) activity,
which may obscure signals from ULDM.
We characterise potential anthropogenic and synanthropic noise sources and examine their influence
on a year-long measurement campaign by AION-10, an upcoming atom interferometer experiment that will be located at the University of Oxford.
We propose a data cleaning framework that identifies and then masks anthropogenic and synanthropic noise.
With this framework, we demonstrate that even in noisy conditions, the sensitivity to ULDM can be restored to within between 10\% and 40\% of an atom shot noise-limited experiment, depending on the specific composition of the anthropogenic and synanthropic noise.
This work provides an important step towards creating robust noise reduction analysis strategies in the pursuit of ULDM detection with atom interferometers.
\end{abstract}


\maketitle

\flushbottom

\section{Introduction}

\begin{figure*}[t]
    \centering
    \includegraphics[width=1.53 \columnwidth]{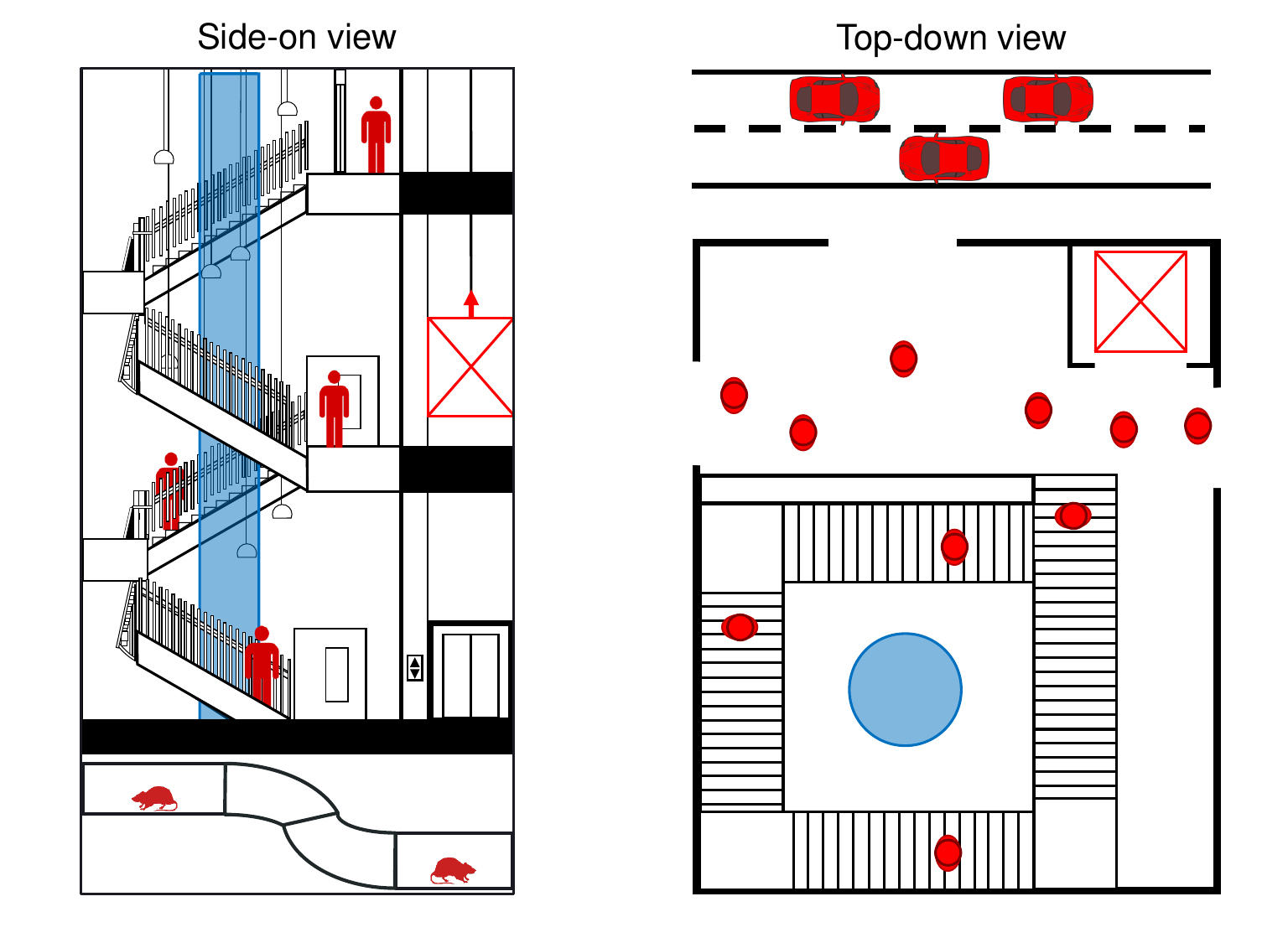}
   \caption{Side-on (left panel) and top-down (right panel) schematic of the Beecroft Building stairwell hosting the AION-10 tower, which is highlighted in blue. 
   Highlighted in red are people moving on the stairs and in the foyer, a nearby passenger lift, traffic on the road outside, and nearby rodents, all of which could  induce a time-dependent phase shift through their gravitational interaction with the atom clouds in the atom interferometers.}
   \label{fig:beecroft}
\end{figure*}

\lettrine{D}{evelopments} in quantum sensing technologies are enabling experiments of unprecedented precision with applications in fundamental physics~\cite{Buchmueller:2022djy}. 
Prominent among these novel sensors are atom interferometers, which have already provided precision measurements of fundamental constants~\cite{Morel, Estey:2014zha, Parker_2018}, tested foundational aspects of quantum mechanics~\cite{Arndt_2014, Bassi:2012bg, Manning:2015cta} and general relativity~\cite{Roura:2018cfg, Zych:2011hu, Xu:2019vlt, Asenbaum_2020}, constrained models of dark energy~\cite{Burrage:2014oza, Hamilton:2015zga, Elder:2016yxm,Sabulsky:2018jma} and `fifth' forces~\cite{Biedermann_2015, Rosi:2017ieh}, and have been proposed as sensors to search for dark matter (DM)~\cite{ Graham:2015ifn,Geraci:2016fva,Arvanitaki:2016fyj,Badurina:2021lwr,Badurina:2021rgt,Badurina:2022ngn,DiPumpo:2022muv} and gravitational waves~\cite{Dimopoulos:2008sv, Dimopoulos:2007cj, Yu:2010ss, Chaibi:2016dze, Graham:2017pmn, Loriani:2018qej,Schubert:2019ycf}.

Light-pulse atom interferometers exploit the wave nature of atoms to split clouds of cold atoms into a superposition of two quantum states that propagate along different paths. 
Laser pulses manipulate the atoms before they coherently recombine and are imaged to measure the phase and contrast acquired by the different arms. 
For a more in-depth discussion of the physical principles, see, for example, Refs.~\cite{Canuel_2006, Hu_2019, Buchmueller:2023nll}. 
The phase measured by atom interferometers is sensitive to changes in timings, atomic structure, and local accelerations making them excellent detectors for DM and gravitational waves. 
Several terrestrial atom interferometer experiments, including AION~\cite{Badurina:2019hst}, ELGAR~\cite{elgar}, MAGIS-100~\cite{MAGIS-100:2021etm}, MIGA~\cite{miga}, and ZAIGA~\cite{Zaiga}, intend to probe unexplored parameter space associated with these phenomena. 
More ambitious projects, such as AEDGE~\cite{aedge} and STE-QUEST~\cite{ste-quest}, aim to take this technology to space for even more sensitive measurements~\cite{Alonso:2022oot}.

The exquisite sensitivity of atom interferometers to changes in the local gravitational potential is well known~\cite{Peters_2001, Kasevich_1992}. 
For instance, lead bricks placed next to atom interferometers have been shown to induce a phase shift on the order of a radian in the interference fringes~\cite{McGuirk2002,Asenbaum_2017}, atom interferometer experiments have been used to measure Newton's gravitational constant~$G_{\mathrm{N}}$~\cite{Fixler_2007,Rosi_2014}, and in the emerging field of quantum cartography, atom interferometers have been used to detect underground cavities and pipes due to the changes in local gravity gradients~\cite{Stray_2022}.

While this sensitivity to the gravitational potential has a myriad of advantages, it could also lead to unwanted signals when searching for DM or gravitational wave signatures.
This is akin to the challenge faced by laser interferometer experiments searching for gravitational waves, where signals from nearby massive objects have been modelled to quantify their contribution to the background noise, see e.g., Refs.~\cite{Creighton_2008, Harms_2019}. 
However, unlike laser interferometers, atom interferometers are most sensitive to signals in the so-called `mid-band frequency range' between 0.1 and 10~Hz. Previous studies have investigated the impact of ground and atmospheric density fluctuations on atom interferometers in the mid-band~\cite{Baker:2012ck, Vetrano:2013qqa, MIGAconsortium:2019efk, Mitchell:2022zbp, Badurina:2022ngn}.
However, the impact of anthropogenic (human sourced) and synanthropic (animal sourced) noise in the mid-band has yet to be explored.

This paper investigates the impact of anthropogenic and synanthropic noise on the upcoming AION-10 atom interferometer experiment~\cite{Badurina:2019hst}. 
AION-10 will have a baseline of approximately $\SI{10}{\meter}$ and operate in a gradiometer configuration, which means that two identical atom interferometers are run simultaneously, launching from the bottom and middle of the baseline. 
Both atom interferometers will operate using the same laser source to  suppress laser phase noise and vibration~\cite{McGuirk2002,Yu2011,Graham:2012sy}.
Single-photon interactions will excite or de-excite the atoms on the $5\mathrm{s}^2\,^1\mathrm{S}_0 \leftrightarrow 5\mathrm{s}5\mathrm{p}\,^3\mathrm{P}_0$ optical clock transition ($\SI{698}{\nano\meter}$ line) in strontium-87~\cite{Hu:2017zft,Hu:2019uey}.
The photon interaction splits the atom into a superposition of momentum states, which creates the two arms of the atom interferometer.
Large momentum transfer (LMT) techniques
apply laser-pulses throughout the sequence to improve the sensitivity by increasing the separation of the interferometer arms~\cite{McGuirk:2000zz,Chiow:2011zz,Graham:2012sy, Rudolph_2020,Wilkason:2022yej}.

Figure~\ref{fig:beecroft} shows a schematic of the stairwell of the Beecroft Building at the University of Oxford's Department of Physics where AION-10 will be hosted. 
As this is an active university building, many sources of anthropogenic and synanthropic noise could potentially surround the tower. 
This includes traceable sources, such as people passing by the tower on the stairs or in the foyer of the building, a passenger lift moving vertically nearby, or vehicles passing on the road outside. 
In addition, there may be sources that are more difficult to trace, such as from random animal transients (RATs), exemplified in Fig.~\ref{fig:beecroft} by small rodents moving near the base of the tower. 
All of these moving masses will cause changes in the local gravitational potential and could lead to unwanted backgrounds.

Searches for ultra-light dark matter (ULDM) with AION-10 will necessitate months or even years of data taking. Previous ULDM sensitivity projections with atom interferometers have neglected the influence of anthropogenic or synanthropic noise. Hence, the primary objective of this work is to characterise potential anthropogenic and synanthropic noise sources in the context of AION-10.

The rest of the paper is structured as follows: section~\ref{sec:FPS} gives a brief review of periodic ULDM signals in atom interferometers and describes the analysis formalism in the frequency-domain for an experiment that has only atom shot noise as a background. Section~\ref{sec:AION-10} characterises the phase contribution from anthropogenic and synanthropic noise sources surrounding AION-10, and describes our simulation of a year-long measurement campaign by the AION-10 experiment.
In section~\ref{sec:freq}, we outline a data-cleaning and mitigation strategy, and present the associated frequency analysis in the presence of anthropogenic and synanthropic noise sources. Finally, in section~\ref{sec:summary}, we provide a summary of our findings and discuss future directions.

\section{Ultra-light dark matter signals}\label{sec:FPS}

For masses below $\sim1\,\mathrm{eV}$, a bosonic ULDM field within our galaxy can be modelled as a set of classical waves~\cite{Hui:2021tkt}.
The coherent oscillation of these ULDM waves can give rise to a variety of time-dependent signals that can be probed with atom interferometers~\cite{Badurina:2019hst, MAGIS-100:2021etm, Badurina:2021rgt}.
This includes the possibility of time-dependent accelerations between atoms in theories of vector and potentially tensor candidates~\cite{Graham:2015ifn,Armaleo_2021},
the time-dependent precession of nuclear spins in the case of pseudoscalars~\cite{Graham:2017ivz}, or time-dependent oscillations of fundamental `constants' in the case of scalar ULDM candidates~\cite{Geraci:2016fva, Arvanitaki:2016fyj, Badurina:2021lwr}.
These signals are in general characterised by a frequency set by the ULDM mass, an amplitude dependent on the local ULDM density and the ULDM mass, 
and a coherence time that depends on the ULDM mass and the ULDM virial velocity in our galaxy.

In this work, we take as our benchmark model a scalar ULDM candidate, $\varphi(t,\mathbf{x})$, but our analysis in sections~\ref{sec:AION-10} and~\ref{sec:freq} can in principle be straightforwardly extended to other ULDM candidates that induce a phase with a regular periodic pattern.
We denote the scalar ULDM mass by~$m_{\varphi}$, and we assume that its characteristic speed is $|\bm{v}|\sim\mathcal{O}(10^{-3})$.\footnote{Here, we follow the convention in the dark matter literature and use natural units with $\hbar=c=1$. However, at other points, we will include explicit factors of $\hbar$.} 
We also assume the scalar ULDM accounts for all of the local dark matter density, so we set $\rho_{\varphi}=0.3~\mathrm{GeV}/\mathrm{cm}^3$~\cite{Read:2014qva}. 
Following Refs.~\cite{Kaplan_2000,Damour:2010rp}, we assume linear interactions
between~$\varphi(t,\mathbf{x})$ and the Standard Model fields,
\begin{equation}
    \mathcal{L} \supset \varphi(t, \bm{x})\sqrt{4\pi G_\mathrm{N}}\left[\frac{d_e}{4e^2}F_{\mu\nu}F^{\mu\nu}-d_{m_e}m_e\bar{\psi}_e\psi_e\right],
\end{equation}
where $e$ is the electric charge ($\alpha=e^2/(4 \pi)$ is the fine-structure constant), $F^{\mu\nu}$ is the Maxwell field strength tensor, $m_e$ is the electron mass, $\psi_e$ is the electron spinor field, and $d_e$ and $d_{m_e}$ are coupling constants parameterising the ULDM interaction with photons and electrons, respectively.\footnote{Quadratic interactions are also viable (see e.g., Refs.~\cite{Stadnik:2015kia, Hees:2018fpg,Bouley:2022eer}) but will not be considered in this work.}

The linear interactions introduce time-dependent corrections to the electromagnetic fine-structure constant and the electron mass~\cite{Stadnik_2014}:
\begin{align}\label{eq:alpha}
    \alpha (t, \bm{x}) &\approx \alpha\left[1+d_e\sqrt{4\pi G_\mathrm{N}}\,\varphi(t, \bm{x})\right], \\
    m_e(t, \bm{x}) &= m_e\left[1+d_{m_e}\sqrt{4\pi G_\mathrm{N}}\,\varphi(t, \bm{x})\right] \label{eq:me}.
\end{align}
Upon modelling the scalar ULDM field as a classical oscillating field,
\begin{equation}\label{eq:ULDMwave}
    \varphi(t, \bm{x}) \approx \frac{\sqrt{2\rho_{\varphi}}}{m_\varphi}\cos(\omega_\varphi t - \bm{k_\varphi}\cdot\bm{x}+\theta),
\end{equation}
where $\omega_\varphi \approx m_\varphi(1+|\bm{v}|^2/2)$, $\bm{k}_\varphi = m_\varphi\bm{v}$ and $\theta$ is a random phase, we see that to leading order in~$|\bm{v}|$, 
the angular frequency of the time-dependent correction to~$\alpha$ and~$m_e$ is set by~$m_{\varphi}$.
Since $\bm{k}_\varphi\cdot \Delta \bm{x}\ll 1$ for all values of~$m_{\varphi}$ accessible to terrestrial atom interferometers, where $\Delta \bm{x}$ is the distance scale over which the atoms propagate, we can safely ignore
the $\bm{k_\varphi}\cdot\bm{x}$ term and focus solely on the time-dependence of the ULDM field. 

Strictly speaking, the amplitude for the oscillating field in eq.~\eqref{eq:ULDMwave} only holds in the regime where the total integration time is longer than the coherence time~\cite{Derevianko:2016vpm}. Otherwise, the amplitude has a stochastic nature that can be modelled with a Rayleigh parameter (see e.g., Refs.~\cite{Foster:2017hbq,Centers:2019dyn,Nakatsuka:2022gaf}). However, the inclusion of a Rayleigh parameter does not impact the mitigation strategies discussed in this work so for simplicity, we use eq.~\eqref{eq:ULDMwave} throughout.

Time-dependent variations in the electron mass or fine structure constant lead to time-dependent transition energies in atoms. For optical electronic transitions of the type
that will be used in AION-10, the time-dependent variation induced by scalar ULDM~is
\begin{align}
    \omega_A(t) &\simeq \omega_A + \overline{\Delta{\omega_A}} \cos(\omega_{\varphi} t + \theta)\,,\\
    \overline{\Delta\omega_A}&=\omega_A\sqrt{4\pi G_\mathrm{N}}\left[d_{m_e}+(2+\xi_A)d_e\right]\frac{\sqrt{2\rho_\varphi}}{m_\varphi}\,,
\end{align}
where~$\omega_A$ is the transition frequency in the absence of~$\varphi(t,\bf{x})$, and~$\xi_A\simeq 0.06$ for the strontium-87 clock transition~\cite{Angstmann:2004zz}. 

\begin{figure*}[!t]
    \centering
    \includegraphics[width=0.89\columnwidth]{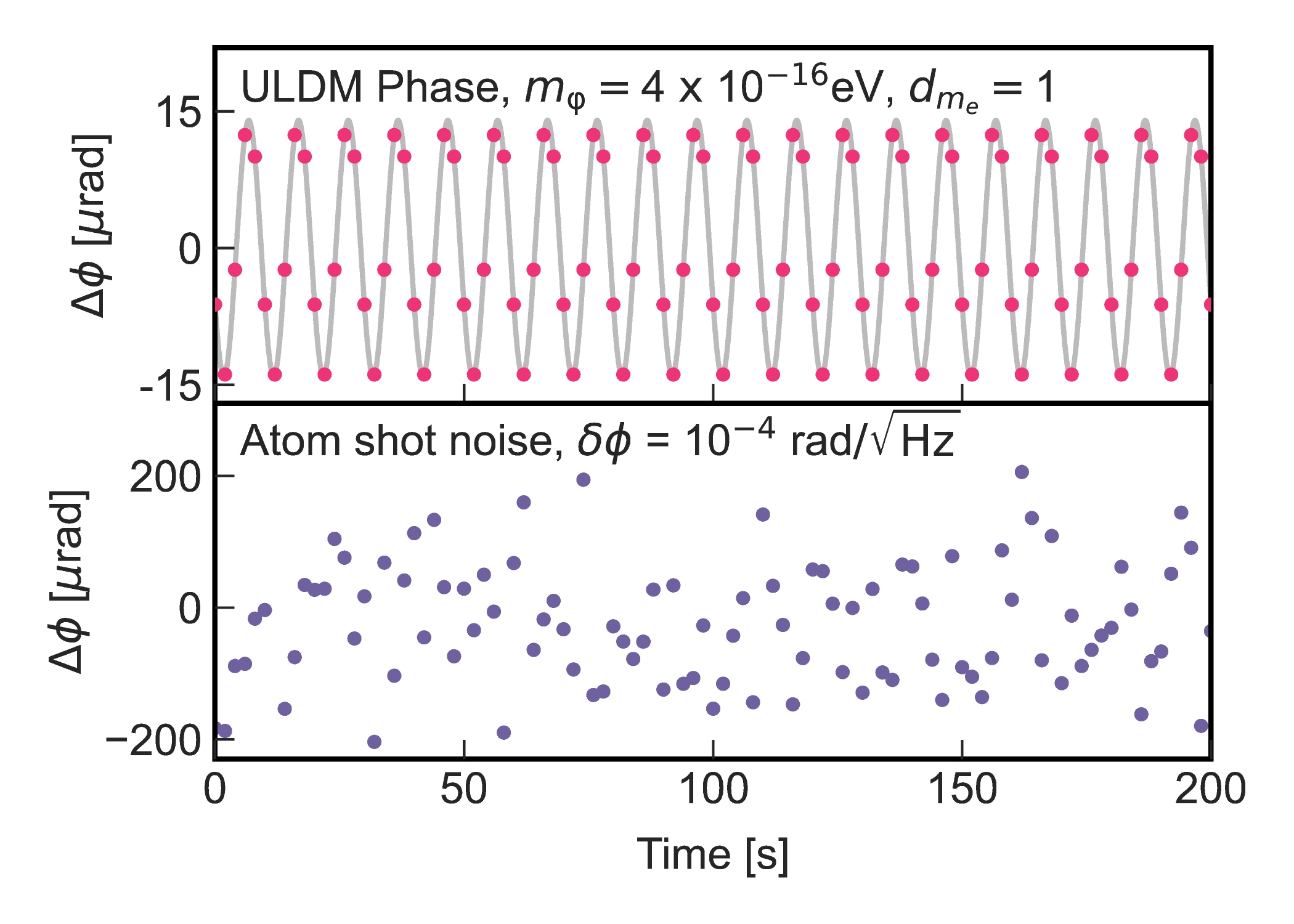}
    \includegraphics[width=\columnwidth]{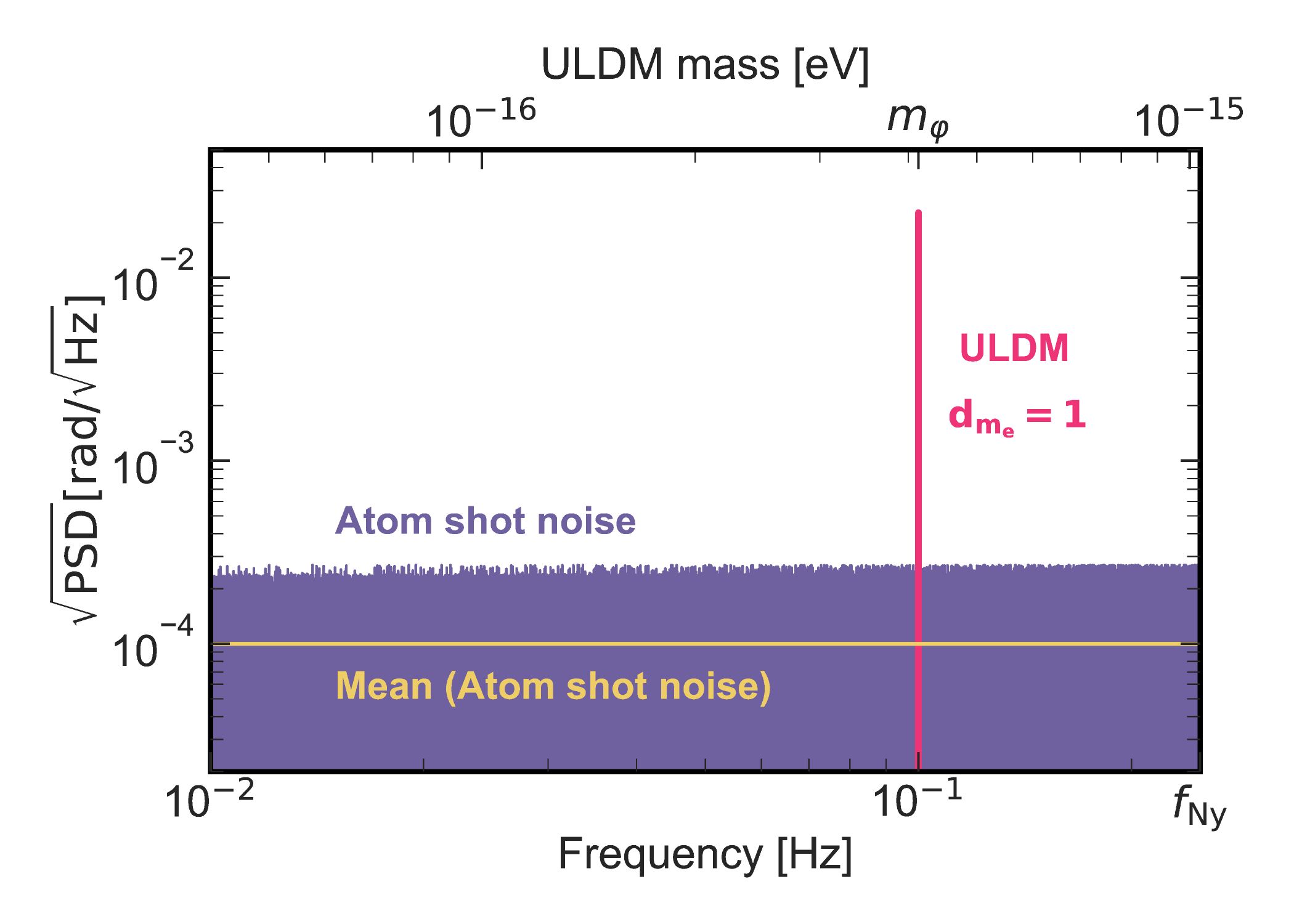}
    \caption{Left: Phase difference between the upper and lower atom interferometers from an oscillating ULDM field (upper panel) with  $m_\varphi \simeq 4\times 10^{-16}$~eV ($f_\varphi = \SI{0.1}{\hertz}$) and $d_{m_e}=1$, and from atom shot noise (ASN, lower panel) assuming a phase noise $\delta\phi = 10^{-4}~\SI{}{\radian\per\sqrt{\hertz}}$. The points represent~50 simulated measurements, assuming $\Delta t = \SI{2}{\second}$. The solid grey line in the upper panel highlights the periodic nature of the differential phase induced by ULDM, compared to the stochastic phase induced by ASN in the lower panel. Right: Square-root of the PSD, $\sqrt{S_k}$, for the ULDM phase signal (pink spike), and the ASN (purple curve) assuming $T_\mathrm{int} =1/3$~yr. The parameters for the ULDM signal and ASN are the same as in the left panels.
    The yellow line shows the root mean square for ASN, which is constant in frequency space, as expected for white noise.}
    \label{fig:DM_SN}
\end{figure*}

The time-dependent nature of the electronic transition induces a phase  in an atom interferometer. For the scalar ULDM search in AION-10, we assume the `broadband' LMT sequence discussed in Refs.~\cite{Arvanitaki:2016fyj, Badurina:2021lwr}, which consists of an initial $\pi/2$-pulse, $(4n-3)$ $\pi$-pulses from alternating directions and a final $\pi/2$-pulse, where~$n$ is the number of LMT kicks.\footnote{A $\pi/2$ $[\pi]$ pulse 
is a pulse of resonant radiation that has a duration of $\pi/(2 \Omega)$ [$\pi/\Omega$], where $\Omega$ is the Rabi frequency.} 
For the gradiometer configuration, the gradiometer phase, i.e.\ the phase difference between the  upper and lower atom interferometers, induced by scalar ULDM for this sequence~is
\begin{equation}\label{eq:DM}
\begin{split}
    \Delta\phi_{\ell} &= 8\,\frac{\overline{\Delta\omega_A}}{\omega_\varphi}\,\frac{\Delta r}{L}\, \sin\left[\frac{\omega_\varphi n L}{2}\right] \\
     &\quad \times \sin\left[\frac{\omega_\varphi\left(T-(n-1)L\right)}{2}\right] \sin\left[\frac{\omega_\varphi T}{2}\right] \\
    & \quad \quad \times\cos\left[\omega_\varphi\left(\frac{2T+L}{2}+\ell \Delta t\right)+\theta\right]\;,
     \end{split}
\end{equation}
where $\Delta r$ is the spatial distance between the two atom interferometers, $L$ is the total baseline, $2T$ is the time between the initial and final $\pi/2$ pulses ($T$ is the `interrogation time'), $\Delta t$ is the temporal separation between successive measurements and $\ell$ is an integer that labels the measurements~\cite{Badurina:2021lwr,Badurina:2022ngn}. For $N$ measurements, $\ell=0$ corresponds to the first measurement and $\ell=N-1$ to the last.  
The amplitude of $\Delta\phi_{\ell}$ is largest when $m_{\varphi}\simeq \left(1.5\,\mathrm{s}/T\right) \times 10^{15}\,\mathrm{eV}$, or in terms of frequency, when $f_{\varphi}=m_{\varphi}/(2\pi)\simeq \left(0.4\,\mathrm{s}/T\right)\,\mathrm{Hz}$. Given $T\sim\mathcal{O}(1)$\,s, AION-10 is most sensitive to scalar ULDM with $m_{\varphi}\sim 10^{-16}$ -- $10^{-14}$\,eV, or equivalently, the frequency range $f_{\varphi} \sim 0.1$ -- $10$\,Hz~\cite{Badurina:2019hst}.

\subsection{Frequency-domain analysis with atom shot noise}\label{sec:FDA}

The pink points in the upper left panel of Fig.~\ref{fig:DM_SN} show the simulated periodic ULDM phase difference for an ULDM field with properties $d_{m_e} = 1$, $d_e = 0$, $m_{\varphi} \simeq 4\times 10^{-16}$~eV ($f_{\varphi}=\SI{0.1}{\hertz}$), and $\theta=\pi/2$. We use the experimental parameters corresponding to the AION-10 (goal) scenario~\cite{Badurina:2019hst}: a baseline $L = \SI{10}{\meter}$, a separation $\Delta r = \SI{5}{\meter}$, and $n=1000$. Assuming that the atom clouds are launched with a speed of $v_\mathrm{l} = \SI{3.86}{\meter\per\second}$, we find $T=\SI{0.73}{\second}$, and we assume~$\Delta t = \SI{2}{\second}$. 

The goal of AION-10 is to reach a phase noise $\delta\phi$ that is dominated by the standard quantum limit, set by atom shot noise (ASN). The ASN scales as $(N_\mathrm{atom})^{-1/2}$, where 
$N_\mathrm{atom}$ is the number of atoms per shot. The purple points in the lower left panel of Fig.~\ref{fig:DM_SN} show simulated phase differences assuming only ASN
with the AION-10 (goal) phase noise of $\delta\phi = 10^{-4}~\SI{}{\radian\per\sqrt{\hertz}}$~\cite{Badurina:2019hst}.

Comparing the upper and lower panels on the left of Fig.~\ref{fig:DM_SN} highlights the challenge of searching for a ULDM signal in the time domain with ASN-limited measurements. The magnitude of the ULDM phase difference is substantially smaller than from ASN, even for $d_{m_e}=1$, which is orders of magnitude larger than current constraints on this parameter~\cite{MICROSCOPE:2022doy}.

Instead, the frequency domain provides a better way to search for the periodic ULDM signal amid the ASN. 
For an {\it uniformly sampling experiment}, the classical periodogram~\cite{schuster1898investigation} is a good estimator of the power spectral density (PSD),
which provides information on the spectral content of the phase time-series.
For a total integration time $T_\mathrm{int}=N \Delta t$, the classical periodogram~is
\begin{equation}
    S_k  = \frac{\Delta t}{N}\left|\sum_{\ell=0}^{N-1} \Delta\phi_\ell \exp\left[-\frac{2 \pi i \ell k}{N}\right]\right|^2,
    \label{eq:PSD}
\end{equation}
where $k=0,\dots,(N-1)$ labels the discrete frequencies, $f_k = k/T_\mathrm{int}$, with a frequency resolution $\Delta f = (T_\mathrm{int})^{-1}$. The maximum non-aliased frequency is given by the Nyquist frequency $f_\mathrm{Ny} = (2\Delta t)^{-1}$. ULDM signals at frequencies higher than $f_\mathrm{Ny}$ are still measurable through aliasing~\cite{Badurina:2023wpk}. Following common usage, we will use the terms periodogram and PSD interchangeably to refer to~$S_k$.

The right panel of Fig.~\ref{fig:DM_SN} shows the square root of the PSD for ULDM (pink `spike') and for ASN (purple curve).
The pink spike occurs at $m_{\varphi}$, the frequency of the phase difference given in eq.~\eqref{eq:DM}, and the spike's amplitude scales with $d_{m_e}$ or $d_e$, and with the integration time~\cite{Badurina:2021lwr}. The stochastic nature of the ASN is reflected in the purple curve, which varies stochastically with frequency. Since the frequency resolution  $\Delta f\sim 10^{-7}\,~\mathrm{Hz}$, when plotted on this scale it appears as a band, so we have also plotted in yellow the root mean square for an ensemble of ASN-only experimental realisations. As expected for white noise, this is constant in frequency space. 

Previous ULDM sensitivity projections for atom interferometers have utilised this frequency-domain framework with the assumption that experimental noise is dominated by ASN
(e.g.,~Refs.~\cite{Arvanitaki:2016fyj,Badurina:2019hst, MAGIS-100:2021etm,Badurina:2021rgt}).\footnote{These projections utilise Bartlett's method when the integration time is greater than the ULDM coherence time, which reduces the ULDM PSD to a spike (single frequency bin). We will also follow this approach.}
While the phase response from rotational effects, magnetic fields, black-body radiation, and from laser phase noise have previously been considered and been shown to be sub-dominant to ASN~\cite{MAGIS-100:2021etm}, the impact of time-dependent anthropogenic
and synanthropic noise on this formalism has yet to be investigated. We will address this in remainder of this paper.

\section{\texorpdfstring{Anthropogenic and \\synanthropic signals}{}}\label{sec:AION-10}

In this section, we describe how we calculate the phase response from anthropogenic and synanthropic sources. 
For each noise source, we evaluate the gradiometer phase; that is, the difference in the phases measured by each atom interferometer: 
\begin{equation}
   \Delta \phi_{\ell} = \phi_{\ell}^{(u)} - \phi_{\ell}^{(l)},
    \label{eq:diffphase}
\end{equation}
where the indices $u, l$ indicate the upper and lower atom interferometer, respectively.

We begin by describing the semi-classical formalism used to calculate the phase for each atom interferometer, where for a quadratic Hamiltonian, Ehrenfest's theorem reduces to Hamilton's equations, so the atoms follow their classical trajectories~\cite{Hogan_2008}.
The classical trajectory in the broadband LMT sequence can be modelled as a Mach-Zehnder sequence in which each pulse transfers a momentum $\pm n\hbar k$, where $k$ is the wave-vector of the laser, to either the upper or lower arm of the atom interferometer.
We assume the atoms launch radially upwards from a spherically symmetric Earth (which defines the $z$-direction) and solve the Euler-Lagrange equations in a gravitational field with acceleration~$g$ and a quadratic term accounting for the Earth's gravity gradient~$\gamma_{zz}$. 
We ignore effects from Coriolis forces as we assume they are corrected with a rotation compensation system, (see, e.g., Ref.~\cite{MAGIS-100:2021etm}).
We then treat the influence of the gravitational potential of people, objects, or RATs nearby, as perturbations on the classical trajectories of the atoms.

The Lagrangian characterising the atoms in the presence of the Earth's gravitational field and a perturbing gravitational potential $V(z, t)$ is therefore
\begin{equation}
    L = m_{A}\left( \frac{1}{2}  \Dot{z}^2 - g z + \frac{1}{2} \gamma_{zz}z^2 - V(z, t) \right),
\end{equation}
where $m_A$ is the mass of the atom.
In general, three contributions to the phase arise: the propagation phase accrued between laser pulses; the laser phase imprinted on the atoms by the laser pulses; and a separation phase when the arms of the interferometer do not overlap at the end of the sequence~\cite{Storey}. 
It is straightforward to calculate the contributions to the phase in the absence of $V(z,t)$ (see, e.g., Refs.~\cite{peters2001high, Bongs_2002,Overstreet_2021}).
The additional phase from the perturbing gravitational potential is found from 
\begin{equation}
\begin{split}
    \delta\phi_{\ell}^{(u,l)} = \frac{m_A}{\hbar} \int_{\ell \Delta t}^{\ell \Delta t + 2T} \mathrm{d}t  &\left[  V\left(z^{(u,l)}_{2}, t\right) \right.
    \\ & \quad - \left. V\left(z^{(u,l)}_{1}, t\right) \right]  \, \,,
    \label{eq:phase1}
\end{split}
\end{equation}
where the indices $1,2$ denote the classical paths travelled by the atoms in each arm of the interferometer in the absence of $V(z,t)$. 

We parameterise the gravitational potential~as
\begin{equation}
    V(z,t) = -\frac{G_\mathrm{N} M}{\sqrt{D(t)^2+\left({z(t)}-h(t)\right)^2}}\,,
    \label{eq:potential}
\end{equation}
where $M$ is the mass of the person, object, or RAT, 
$D(t)$ and $h(t)$ are the horizontal and vertical distances from the atoms, 
and $z(t)$ is the vertical height of the atoms during the sequence. All vertical distances are defined relative to the atom launch position in the lower atom interferometer.

The semi-classical formalism that we have used is exact for a quadratic potential.
Potentials that are not quadratic, which includes the gravitational potential, will give non-classical corrections.
The size of the correction can be estimated by considering
the leading correction to $\partial_t\langle\hat{p}\rangle$, which is  $\tfrac{1}{2} (\partial_r^3V(r))\Delta x^2$,
where $\langle\hat{p}\rangle$ is the momentum operator expectation value and $\Delta x^2$ is the variance of atom wave-packet size~\cite{Hogan_2008, Bertoldi:2018zga, Ufrecht:2020vcr, Overstreet_2021}.
For typical AION-10 parameters, we find the leading non-classical correction is $ \sim 10^{-16}g \left( M/80\,\mathrm{kg} \right)   \left( 2\,\mathrm{m}/r \right)^4 \left(\Delta x^2/ 1\,\mathrm{mm}^2 \right)$.
Given that 10\,m atom interferometers operate at a sensitivity $\sim10^{-12}g$ (see, e.g., Ref.~\cite{Overstreet:2021hea}), the semi-classical formalism is sufficiently accurate for the gravitational potentials that we consider.

\subsection{Characterising background-induced phases}

\begin{figure}[t!]
    \centering
    \includegraphics[width=0.9\columnwidth]{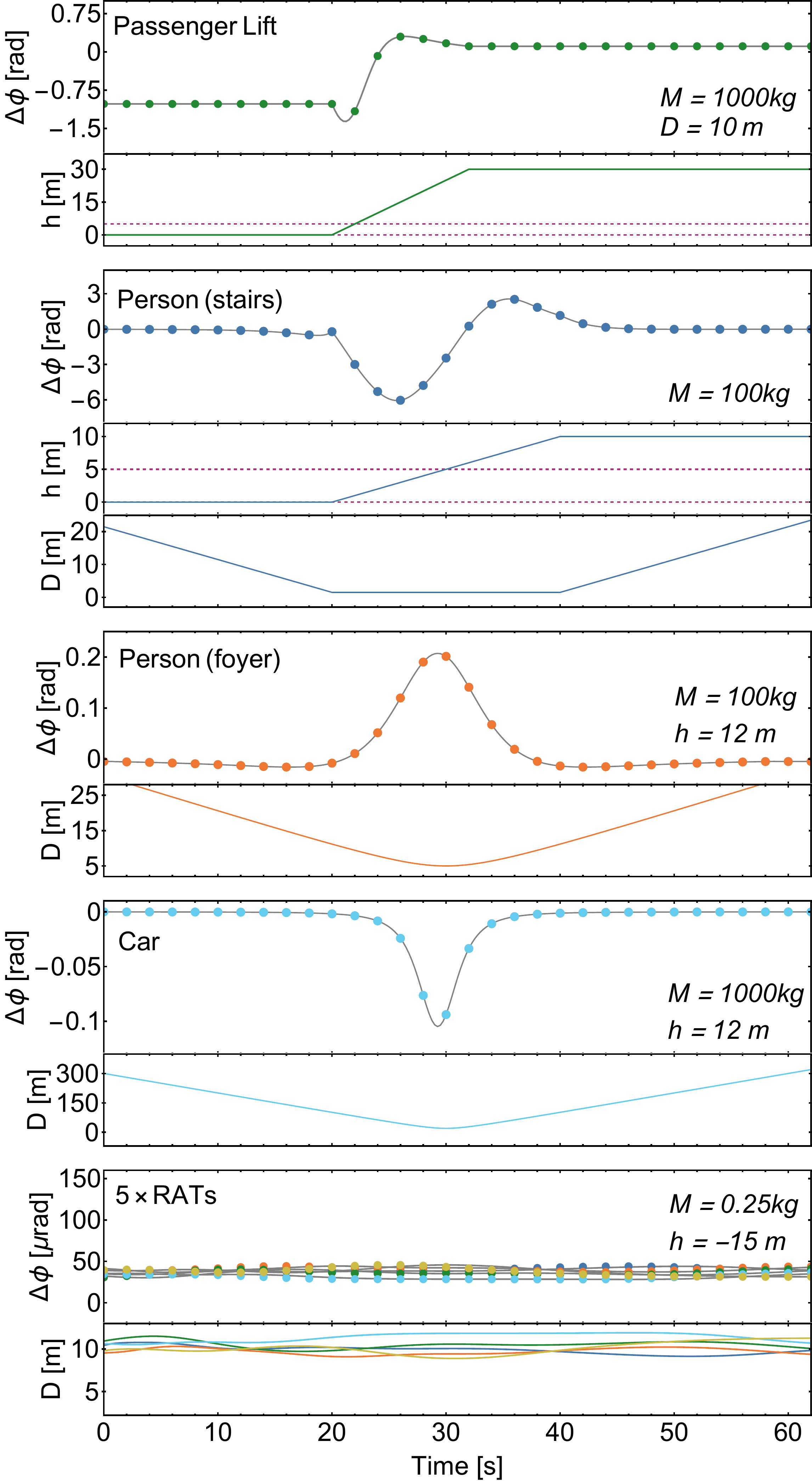}
    \caption{Upper part of each panel shows the simulated time-dependent phases induced by the gravitational potential of anthropogenic and synanthropic noise sources. 
    From top-to-bottom, we consider: a passenger lift (elevator), a person moving on stairs and in the foyer, a car, and five random animal transients (RATs).
    The time-dependence of the distance parameters are plotted in the lower part of each panel. 
    The position of the two atom interferometer launch points are shown as dashed lines at $h=\SI{0}{\meter}$ and $h=\SI{5}{\meter}$.
        The largest differential phase is induced by the person on the stairs, due to their  proximity to the AION-10 tower.}
    \label{fig:PS}
\end{figure}

Having defined the semi-classical formalism, we are now in a position to determine the phase response from anthropogenic and synanthropic sources that could impact AION-10.
Figure~\ref{fig:PS} shows the simulated phase response from a single incidence of five possible noise sources: a passenger lift; a person climbing stairs next to the tower; a person walking across the foyer next to the top of the tower; a car travelling past on a nearby road; and a small animal moving close to the base of the tower.\footnote{The gravity gradient gives a constant contribution $\Delta \phi_{\ell}= n  k \gamma_{zz} \delta r T^2$, which we do not include when plotting $\Delta \phi_{\ell}$, so as to clarify the impact of the noise sources. Laser-pulse sequences can be engineered to minimise the gravity gradient~\cite{Roura:2015xsa,PhysRevLett.119.253201}, but we do not investigate these schemes in this work.} The upper part of each panel shows the time-dependence of the gradiometer phase, while the lower part shows the most relevant time-dependent distance parameter that enters eq.~\eqref{eq:potential}.

The first two examples consider scenarios where the dominant effect occurs from vertical movement parallel to the tower.
The upper panel models a passenger lift that starts from rest at the base of the AION-10 tower, moves up past the top of tower at a constant speed of $v=\SI{2.5}{\meter\per\second}$, and then stops. 
The behaviour of the phase mirrors this by initially recording a constant value, changing as the lift moves, before finally plateauing to a constant value.
 We model the lift with a constant mass and assume that there is no counterweight mechanism.
 The second panel models the response from a person walking at constant speed $v=\SI{1}{\meter\per\second}$ to the base of the tower, then up the stairs next to the tower at constant speed $v=\SI{0.5}{\meter\per\second}$, before finally walking away.
The phase is initially small, changes significantly as the person climbs the stairs, and tends to zero after the person walks away.
Were the lift to move in the opposite direction, or a person to walk down the stairs, the phase curve would be reversed.
 
Next, we consider two scenarios where the dominant effect occurs predominantly from horizontal movement relative to the tower: a person walking at constant speed $v=\SI{1}{\meter\per\second}$ along the foyer adjacent to the stairwell that hosts AION-10; and a car travelling at constant speed $v=\SI{10}{\meter\per\second}$ on a straight road.  The shape of the phase response is similar in both cases, although there is a relative sign change and the shape is narrower for the car as it moves faster.

Finally, the simulated phase response from five small animals (RATs) moving quasi-randomly under the base of the AION-10 tower is shown in the bottom panel of Fig.~\ref{fig:PS}.
The phase response of the small animal induces $\mathcal{O}(50)\,\mu\mathrm{rad}$ `ripples' in the time series, which are significantly smaller than the phase from the other sources. However, the total induced response is similar in magnitude to atom shot noise ($\mathcal{O}(250)\,\mu\mathrm{rad}$ cf.\ Fig.~\ref{fig:DM_SN}). As such, we include RATs in our analysis to represent a class of noise sources that give a contribution somewhat similar to atom shot noise, and which may be difficult to mitigate against in analysis strategies.

It is relatively straightforward to gain intuition for the results in Fig.~\ref{fig:PS} by considering a limiting case of the phase response calculation where analytic results can be obtained. 
To that end, for this discussion we will ignore the effect of the Earth's gravity gradient on the classical trajectory of the atoms since it results in more complicated expressions with little gain in intuition.
First, by considering eqs.~\eqref{eq:phase1} and~\eqref{eq:potential}, it is trivial to observe that for an object of constant mass $M$, $V(z,t)\propto M$ so $\delta \phi_{\ell}^{(u,l)}\propto M$ and hence $\Delta \phi_{\ell}\propto M$, i.e.\ the phase response is simply linear in $M$.

\begin{figure*}[t!]
    \centering
      \includegraphics[width=1.99\columnwidth]{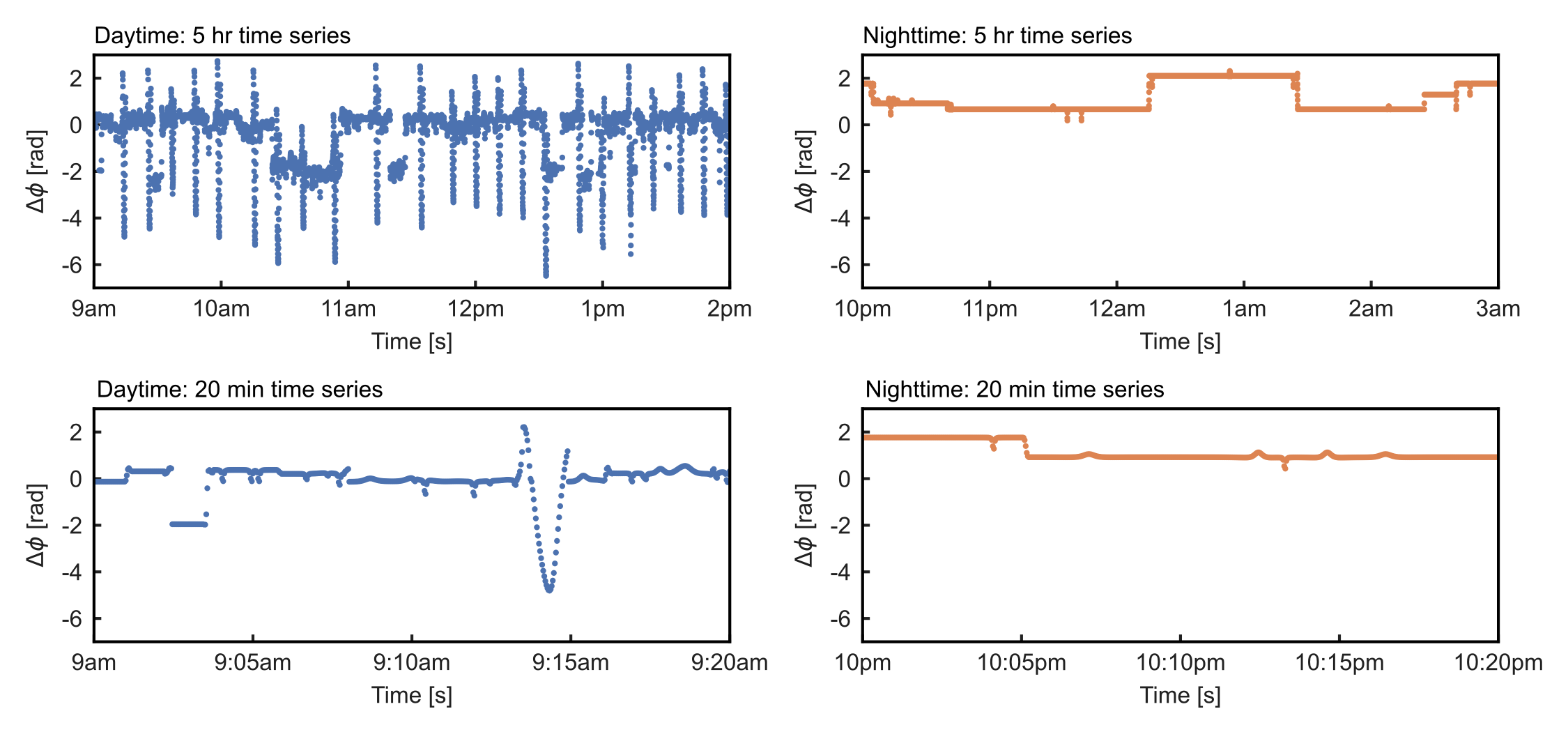}
    \caption{Simulated differential phase induced by a ULDM signal, atom shot noise, and anthropogenic and synanthropic noise sources.
    The left and right panels show a simulated time series during the daytime and nighttime, respectively. The lower panels focus on the first 20 minutes of the time series.
    The phase is dominated by the anthropogenic noise sources.}
    \label{fig:totPS}
\end{figure*}

Next, consider a stationary object of mass $M$ in the regime where $D, h \gg z(t)$. 
Expanding the potential, the first non-zero contribution to $\Delta \phi_{\ell}$ arises from $V(z,t)=-\tfrac{1}{2} G_{\mathrm{N}} M z^2 (D^2-2h^2)/(D^2+h^2)^{5/2}$, and gives a phase 
\begin{equation} \label{eq:scaling} 
\Delta \phi_{\ell}\simeq - n k T^2 \,G_{\mathrm{N}} M \Delta r\, \frac{D^2-2 h^2}{(D^2+h^2)^{5/2}} \;.
\end{equation}
In the limit $D\gg h$, we see that the gradiometer phase is negative and scales with an inverse cube law with distance from the atom interferometers. 
This power-law scaling with distance relative to the linear scaling with mass is the reason that a single person walking on the stairs next to the tower induces a much larger phase than the more massive passenger lift that is further away.
We also see from eq.~\eqref{eq:scaling} that we get a sign change when $D\lesssim h$, which is the behaviour we see in Fig.~\ref{fig:PS} for the person (foyer) relative to the car: 
both have the same value of $h$ but the minimum $D$ value is $\SI{5}{m}$ for the person (foyer) and $\SI{20}{m}$ for the car. For the relatively small values of $D$ and $h$ used in Fig.~\ref{fig:PS}, next-order corrections to $\Delta \phi_{\ell}$ modify the relation such that the sign changes at $D\sim h$ rather than $D=\sqrt{2} h$.

\subsection{Simulated daytime and nighttime time series \label{sec:simtimeseries}}

Over an extended period of data taking, many anthropogenic and synanthropic sources will contribute to the total phase readout.
 Figure~\ref{fig:totPS} shows a simulation of extended data-taking periods: 
 the left panels show the time series from a busy period during the day, 
 while the right panels show the time series from a quiet period at night when the occupation levels in the building are significantly lower. The upper panels show a five~hour time series, while the lower panels zoom in on the first 20~minutes.

We make the following assumptions for the daytime simulation. 
Firstly, we assume the lift moves randomly between seven floors within the building at a rate of $20$ movements per hour between 9-10am and 12.30-2pm, five movements per hour between 10-12.30pm, and we assume the mass varies randomly between $1000$ and \SI{2000}{kg} to account for a different number of people during each movement;
Secondly, we assume that five people per hour use the stairs next to the tower and that each person has a mass ranging from $50$ to \SI{100}{kg}.
Thirdly, we model $100$ people per hour walking across the foyer or on floors above the tower between 9-10am and 12.30-2pm, and 25 people per hour between 10-12.30pm.
Finally, we assume $60$ vehicles pass per hour, with a mass randomly selected between $1000$ and \SI{5000}{kg}.

For the nighttime simulation, our assumptions are similar but the rates are reduced.
Firstly, the lift moves once per hour at a random time throughout the night. 
Secondly, owing to the large induced phase from the people on the stairs, we assume that the stairs next to the tower are closed at night.
Thirdly, we model up to ten people in the foyer and up to ten cars passing throughout the night.
In both the daytime and nighttime simulations, we include the contribution from atom shot noise, RATs and ULDM, but their contribution is not visible on the scale in Fig.~\ref{fig:totPS}.

In all panels in Fig.~\ref{fig:totPS}, we see that the time series is characterised by large transients on short timescales, $\mathcal{O}(1)\,\mathrm{min}$, that often resemble spikes or oscillations, together with longer periods (minutes or even hours) where the time series is relatively flat.
The spiked features can be understood with reference to Fig.~\ref{fig:PS}: the change in one plateau to another results from a movement of the lift; the transient down and up behaviour follows from a person on the stairs; and the Gaussian shapes are people on the foyer or vehicles passing by.
Unsurprisingly, the daytime simulated time series has significantly more variation compared to the nighttime simulation.

\section{\texorpdfstring{Analysis strategy and \\sensitivity projections}{}}\label{sec:freq}

Removing noise from time series data is a problem encountered in many fields (see, e.g., Refs.~\cite{Deeming1975, Carbonell1992, huppert2009homer, Feuerstein2009, PATTON2012, Vaughan:2013bca, Desherevskii2017}).
While a menagerie of techniques are known, we utilise an approach that is aligned with our intended outcome:
to achieve sensitivity to ULDM that approaches an atom shot noise-limited experiment.
Our discussion in this section has two goals. 
As we are working with simulated data, which contains a representative rather than fully comprehensive set of possible noise sources,
our first goal is to provide a high-level discussion of the data cleaning strategy. 
Our expectation is that the methods will be extended upon application to real data but the principles will remain.
Our second goal is to address the potential impact of anthropogenic and synanthropic noise sources on ULDM sensitivity projections.

To proceed, we exploit information about the ULDM signal and noise sources.
Firstly, we have shown in section~\ref{sec:AION-10} that anthropogenic or synanthropic noise sources induce time-dependent contributions to the differential phase on $\mathcal{O}(1)\,\mathrm{min}$ timescales, separated by longer periods of stability. 

Secondly, any noise-reduction procedure must ensure that distortions to the periodic structure in the ULDM signal are minimised. This is because the amplitude of the ULDM signal in the time-domain is much smaller than atom shot noise (cf.\ Fig.~\ref{fig:DM_SN}), so our analysis relies on recovering the periodic signal through a frequency domain analysis. 

Thirdly, the sensitivity to the coupling constants parameterising the ULDM interactions with photons or electrons scales with the total integration time as $(T_{\rm{int}})^{-1/4}$ when the integration time is greater than the ULDM coherence time, or $(T_{\rm{int}})^{-1/2}$ in the opposite regime~\cite{Badurina:2021lwr}.
This means that removing parts of the time series and thereby reducing $T_{\rm{int}}$ results in a relatively small loss in sensitivity.
In addition, an even more drastic choice of restricting to nighttime running (i.e., to 8\,hours instead of 24\,hours), when the time series is significantly cleaner, reduces sensitivity to most of the mid-band frequency range by only $30\%$, rather than a factor of three.\footnote{We have not considered the impact of anthropogenically induced seismic noise but minimising this source also motivates running at night.} 

All of these considerations motivate nighttime running combined with the application of masks to remove the anthropogenic- or synanthropic-induced transients from the time series. An immediate implication of using masks is that we have to analyse a time series with gaps. 
Equivalently, this can be viewed as a non-uniformly sampled time series.
Analysis of `gapped' or non-uniformly sampled time series is common in astronomy, since ground-based observations occur at night and clouds or weather can prevent observations for short periods. Therefore, rather than using the classical periodogram, we instead use an estimator of the PSD that is commonly used in the astrophysics community and which is designed for non-uniformly sample data: the Lomb-Scargle periodogram~\cite{lomb1976least,scargle1982studies}.

In the rest of this section, we describe our data cleaning strategy,
review properties of the Lomb-Scargle periodogram, and examine the PSD derived from time series encompassing both anthropogenic and synanthropic noise sources, both with and without applying the data cleaning techniques.
Finally, we present the impact of anthropogenic and synanthropic noise sources on ULDM sensitivity projections after the data cleaning strategies have been employed.

\begin{figure*}[!t]
    \centering
    \includegraphics[width=1.99 \columnwidth]{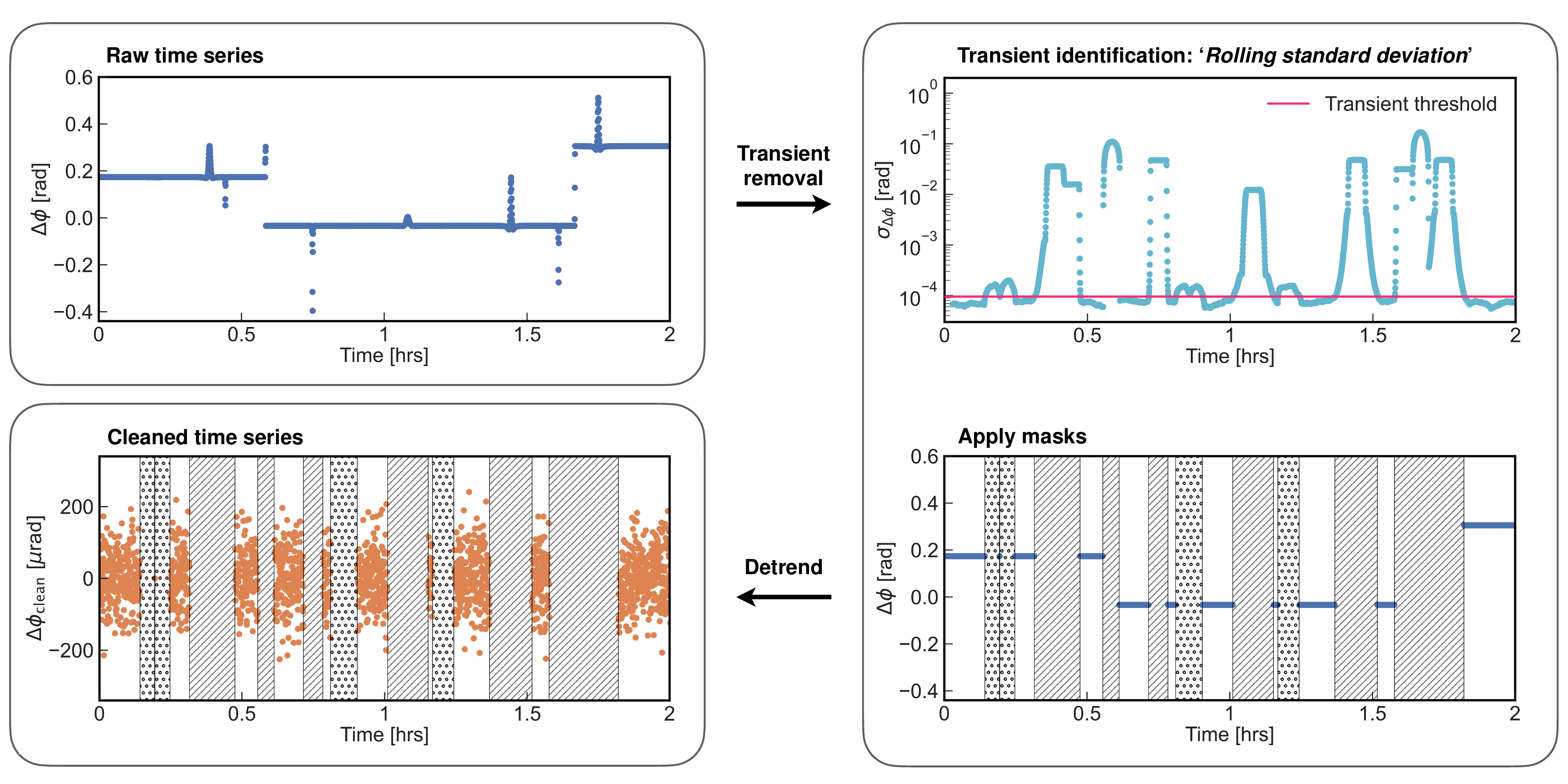}
    \caption{Our data cleaning approach to eliminate transient noise in a simulated time series. 
    The upper left panel shows the raw time series containing transients in the phase induced by anthropogenic noise and (invisible on this scale) changes induced by synanthropic noise.
    Transients are identified by calculating $\sigma_{\Delta\phi}$, the rolling standard deviation (upper right).     
    Events that surpass the transient detection threshold, indicated by the magenta line, are masked, as indicated in the lower panels: 
    diagonal lined hatching represents masking applied to anthropogenic noise while the dotted hatching represents masked synanthropic noise.
    The bottom left panel shows the result of detrending the masked time series. The cleaned time series is now distributed around a common baseline and is ready for analysis with the Lomb-Scargle periodogram.}
    \label{fig:mask}
\end{figure*}

\subsection{Cleaning the time series}

Our data cleaning approach is summarised in Fig.~\ref{fig:mask}. We start with the raw time series shown in the upper left panel of Fig.~\ref{fig:mask}, which has been  created to have multiple transients spread throughout the time period. The first stage is to identify and remove  transients from the time series. Several approaches have been developed to identify transients, including moving average techniques, median filtering, change point detection and machine learning algorithms (see, e.g.,~Ref.~\cite{nielsen2019practical}).
Moving averages are generally more effective at detecting short-term, localised transients of the type that appear in our raw time series~\cite{shumway2000time}.
Therefore, we adopt a moving average of the standard deviation algorithm, which for simplicity, we refer to as the rolling standard deviation.

We define the rolling standard deviation for the $\ell^{\mathrm{th}}$ phase in the series as
\begin{equation}
    \sigma_{\Delta \phi_{\ell}} = \sqrt{\frac{1}{N-1}\sum_{i=\ell-(N-1)/2}^{\ell+(N-1)/2} \left(\Delta\phi_i - \overline{\Delta\phi}_\ell\right)^2\,},
\end{equation}
where $N$ is an odd integer that defines the window size and $\overline{\Delta\phi_\ell}$ is the rolling mean, defined as
\begin{equation}
     \overline{\Delta\phi_\ell} = \frac{1}{N}\sum_{i=\ell-(N-1)/2}^{\ell+(N-1)/2} \Delta\phi_i\;.
\end{equation}
The choice of the window size is important.
While a larger $N$ can provide more stability in the rolling standard deviation, a value that is too large may miss localised transients.
We found that a step size of $N=101$ strikes a good balance between these two considerations. 

The upper right panel of Fig.~\ref{fig:mask} shows the rolling standard deviation for the raw time series shown in the upper left panel.
Transients are identified when they cross a transient threshold, shown by the solid pink line. Our threshold is set as the mean-plus-five-standard-deviations of an ASN-limited experiment.
Transients caused by anthropogenic noise are orders of magnitude above the threshold so are easily identified. In contrast, transients caused by synanthropic noise appear just above threshold. Events that surpass the rolling standard deviation are masked.
For the time series in Fig.~\ref{fig:mask}, the masked regions are represented by hatched patterns in the lower panels, where diagonal hatching represents anthropogenic noise and dotted hatching represents synanthropic noise. 

The final stage involves detrending the masked time series to eliminate any long-term drifts and establish a common baseline. We achieve this by fitting a linear regression model to each region of the time series between the masks. Subtracting the linear fit from each region effectively detrends the time series. This detrending approach works well with our simulated data. However, it may require modifications when applied to real data, where quadratic trends or other longer-term drift effects could be more pronounced and require special consideration.
The cleaned time series looks similar to the ASN-only time series shown in Fig.~\ref{fig:DM_SN}, but with gaps at the regions where the masks have been applied.

\subsection{Lomb-Scargle periodogram}

We employ the Lomb-Scargle periodogram to analyse the spectral features of the cleaned time series. 
The Lomb-Scargle periodogram is usually defined as
\begin{equation}\label{eq:LS}
\begin{split}
    S^{\mathrm{LS}}_k &= \frac{1}{2}
    \frac{\left(\sum_\ell \Delta\phi_\ell \cos{\left(2\pi f_k\left[t_\ell-\tau\right]\right)}\right)^2}{\sum_\ell \cos^2(2\pi f_k[t_\ell - \tau])} \\
    &\qquad \quad + \frac{1}{2} \frac{\left(\sum_\ell \Delta\phi_\ell \sin(2\pi f_k [t_\ell-\tau])\right)^2}{\sum_\ell \sin^2(2\pi f_k [t_\ell-\tau])} \,,
    \end{split}
\end{equation}
where $\ell$, as before, is an integer running over the values from $0$ to $N-1$,~$t_{\ell}$ denotes the time of the $\ell^{\mathrm{th}}$ measurement, $f_k$~denotes discrete frequencies, and~$\tau$ is given for each frequency $f_k$ by
\begin{equation}
    \tau = \frac{1}{4\pi f_k}\tan^{-1}\left(\frac{\sum_\ell \sin(4\pi f_k t_\ell)}{\sum_\ell \cos(4\pi f_k t_\ell)}\right)\;.
\end{equation}
See Ref.~\cite{VanderPlas_2018} for a more comprehensive discussion.

 As defined here, $S^{\mathrm{LS}}_k$ has units of $\mathrm{rad}^2$, which is different  to our definition of the classical periodogram, $S_k$, which has units $\mathrm{rad}^2/\mathrm{Hz}$. The Lomb-Scargle periodogram allows for differing values of $\Delta t$ throughout the time series, so dropping the $\Delta t$ prefactor that was included in eq.~\eqref{eq:PSD} avoids ambiguity.
 In our simulation, however, we assume a structured observation regime in which we fix $\Delta t =\SI{2}{\second}$ for all observations. Therefore, to make a closer connection with the earlier results, we show $\mathrm{PSD}_{\rm{LS}}\equiv\Delta t\,  S^{\mathrm{LS}}_k$ in our figures.

 In practical terms, we calculate $S^{\mathrm{LS}}_k$ using the \texttt{astropy} package~\cite{astropy:2022}, with the `psd' normalisation and `fast' method, which utilises the fast Fourier transform. This formalism fits a sinusoid to the data at each frequency with a higher power in the periodogram being associated with a better fit. It uses the floating-mean model to fit for the mean of data and uses only one Fourier term in the fit.
We set `$\text{samples\_per\_peak}=1$'  such that $f_k=k/T_{\rm{int}}$, which implies that our frequency grid and Nyquist frequency are the same for both the Lomb-Scargle and classical periodograms.
 
\subsection{Frequency-domain analysis with anthropogenic and synanthropic noise \label{sec:freqenvnoise}}

\begin{figure}[!t]
    \centering
    \includegraphics[width=\columnwidth]{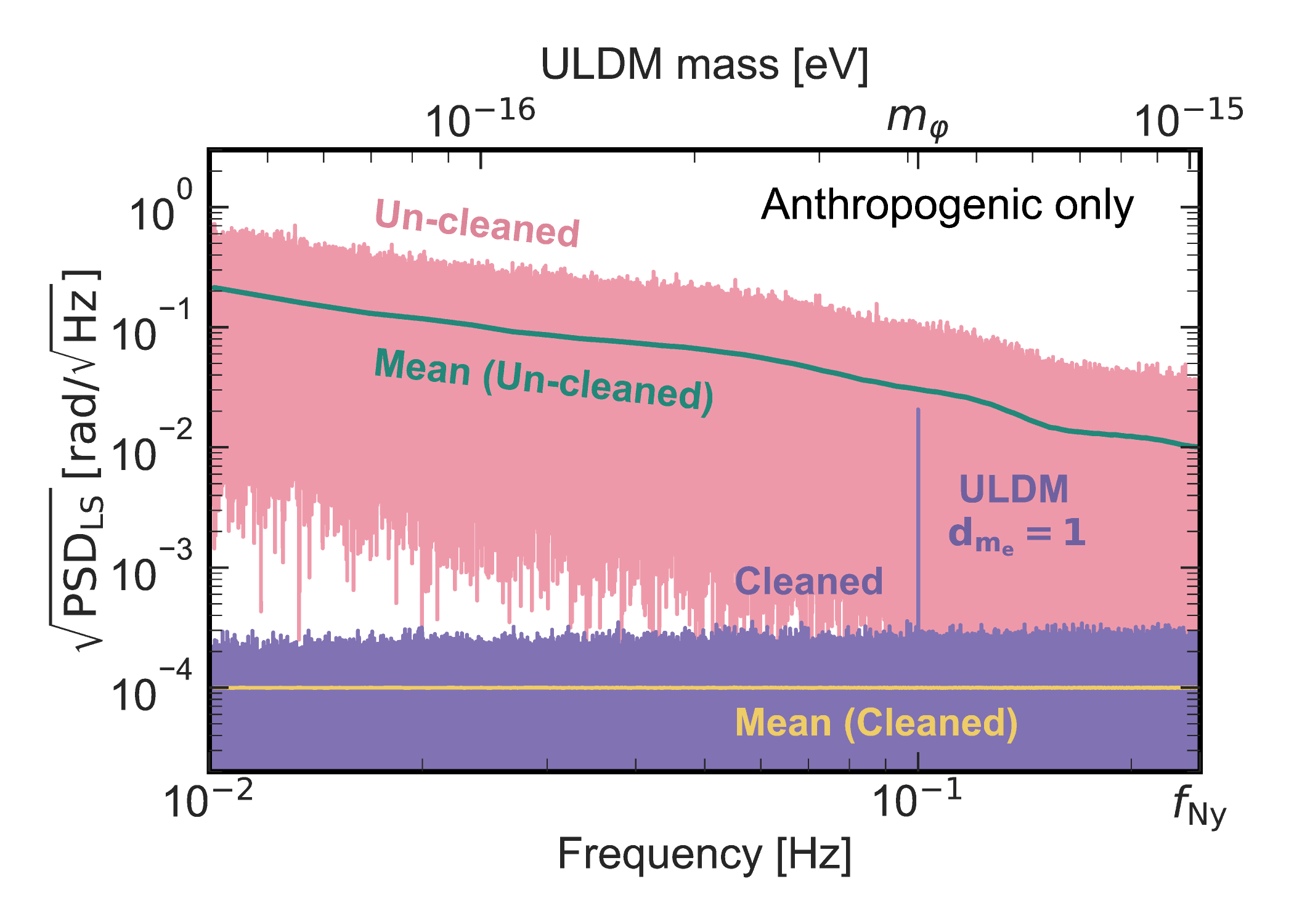}
    \includegraphics[width=\columnwidth]{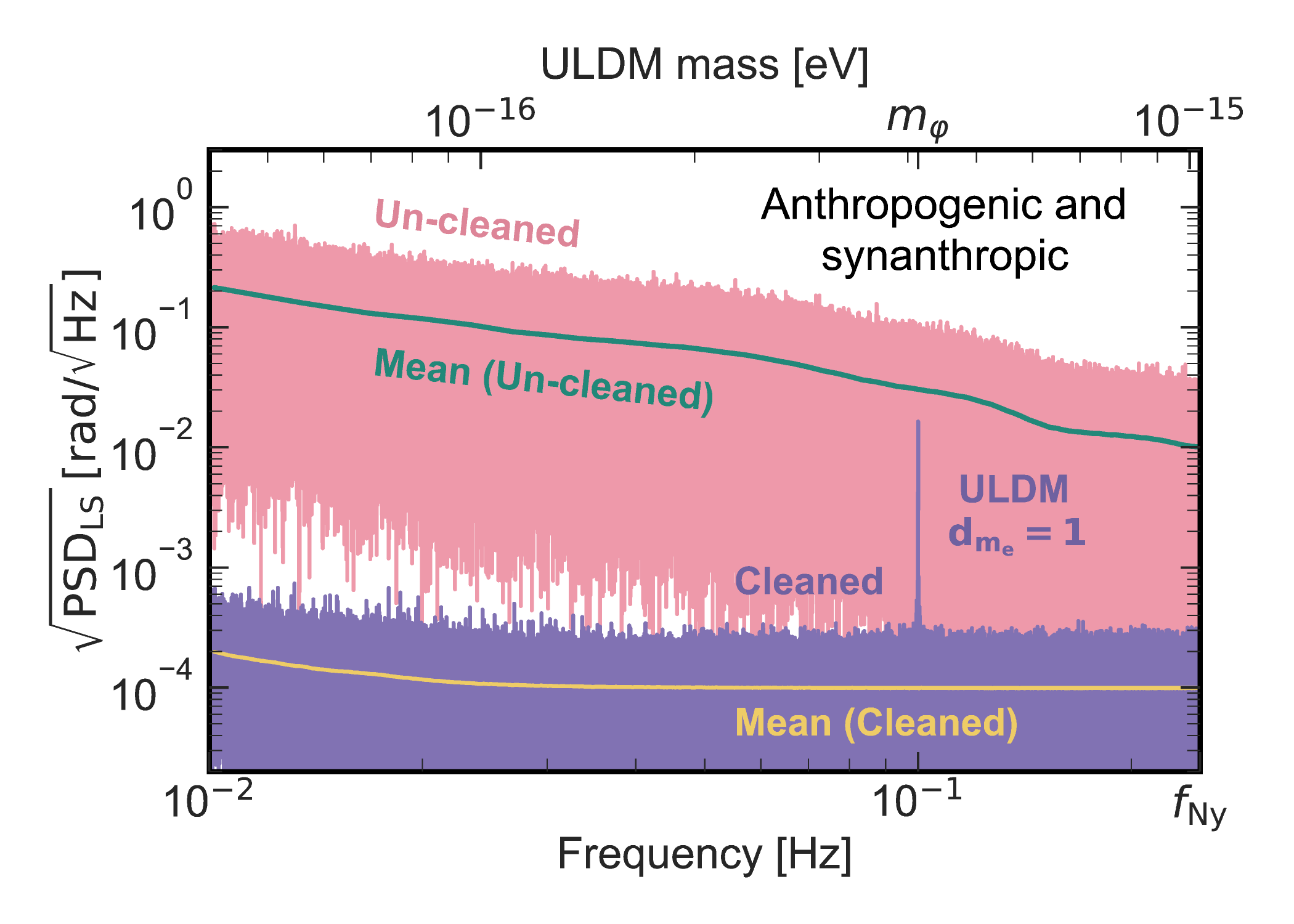}
    \caption{ Upper panel: the square root of the Lomb-Scargle PSD for a time series that includes contributions from a ULDM field, ASN, and anthropogenic noise associated with nighttime running. 
Lower panel: the PSD is extended to include the contribution from synanthropic noise in addition to the aforementioned components.
The pink line shows the PSD of the raw time series without any cleaning strategies applied, assuming nighttime running for 365~days. The purple line shows the PSD of the time series after the implementation of our cleaning strategy.
The yellow (green) lines show the root mean square with (without) cleaning.
The cleaned PSDs closely resemble the ASN-only case indicating that the data cleaning method has successfully identified and removed anthropogenic and synanthropic transients. 
    }
    \label{fig:unCutPSD}
\end{figure}

The two panels in Fig.~\ref{fig:unCutPSD} show the square root of the Lomb-Scargle PSD for a simulated time series with a ULDM field with parameters $m_{\varphi}\simeq 4\times 10^{-16}\,\mathrm{eV}$ ($f_{\varphi}=0.1\,\mathrm{Hz}$) and $d_{m_e}=1$, ASN with phase noise $\delta \phi = 10^{-4}\,\mathrm{rad}/\sqrt{\mathrm{Hz}}$ and anthropogenic noise (upper panel), and with both anthropogenic and synanthropic noise (lower panel). The simulated time series assumes that data taking occurs during an eight-hour period at nighttime each day over the course of one year, for a total (raw) integration time of $1/3~\mathrm{yr}$. 

The assumptions made in generating the nighttime anthropogenic noise are discussed in section~\ref{sec:simtimeseries}. The synanthropic noise in the time series that enters the lower panel of Fig.~\ref{fig:unCutPSD} is generated with the following assumptions: 
we simulate five RATs with a mass $\SI{0.25}{\kilogram}$ running on pseudo-random paths underneath the AION-10 detector. 
Each RAT on average runs on a pseudo-random path 300 times per night.
The paths are randomly generated on a $\SI{10}{\meter}$ $\times$ $\SI{10}{\meter}$ horizontal grid that is $\SI{15}{\meter}$ below the atom launch position in the lower atom interferometer.  

In both panels, the pink curve shows the Lomb-Scargle PSD obtained from the raw time series without any cleaning strategies applied (when plotted on this scale, it appears as a large band). The green line shows the root mean square. By comparing the upper and lower panels, we see that the pink and green lines are similar, showing that, as we would expect for both scenarios, the uncleaned PSD is dominated by anthropogenic noise. We also observe that the mean and standard deviation of the uncleaned PSD is at least two orders of magnitude larger than the ASN-only example shown in Fig.~\ref{fig:DM_SN}. The result is that the ULDM spike with the stated parameters cannot be observed with the uncleaned PSD.

We next consider the purple curves in both panels of Fig.~\ref{fig:unCutPSD}, which are the Lomb-Scargle PSDs obtained after the application of our data cleaning approach. 
The yellow lines show the root mean square.
In the upper panel, we observe a cleaned PSD that closely resembles the ASN-only case:
 the mean and standard deviation are both $10^{-4}\,\SI{}{\radian\per\sqrt{\hertz}}$ and are independent of frequency for the frequency range plotted. 
 This indicates that the data cleaning method has successfully identified and removed anthropogenic noise.
 As a result, atom shot-noise limited sensitivity is achieved, albeit with a reduction in the integration time owing to the presence of masks. The masks have reduced the number of points in the raw time series and therefore the effective integration time by $16.3\%$. Furthermore, the ULDM spike is clearly identified in the cleaned PSD at 0.1~Hz, which indicates minimal distortion to the ULDM signal. The amplitude of the ULDM spike in the upper panel is $\SI{2.07e-2}{\radian\per\sqrt{\hertz}}$, compared to $\SI{2.27e-2}{\radian\per\sqrt{\hertz}}$ in Fig.~\ref{fig:DM_SN}. Since $S_k \propto T_\mathrm{int}$~\cite{Badurina:2021lwr}, the reduction in power in the ULDM spike follows as a result of masking part of the raw time series.

In the lower panel, where we account for synanthropic noise alongside anthropogenic noise, the cleaned PSD exhibits a close resemblance to the ASN-only case for frequencies above approximately $0.04\,\mathrm{Hz}$. However, at lower frequencies, the PSD rises due to the presence of synanthropic noise. Additionally, the inclusion of synanthropic noise leads to more of the time series being masked, resulting in a further reduction of the ULDM spike amplitude to $\SI{1.67e-2}{\radian\per\sqrt{\hertz}}$, as 45.7\% of the points are masked in this instance.

\subsection{Impact on ULDM sensitivity projections}

\begin{figure}
    \centering
    \includegraphics[width=0.93 \columnwidth]{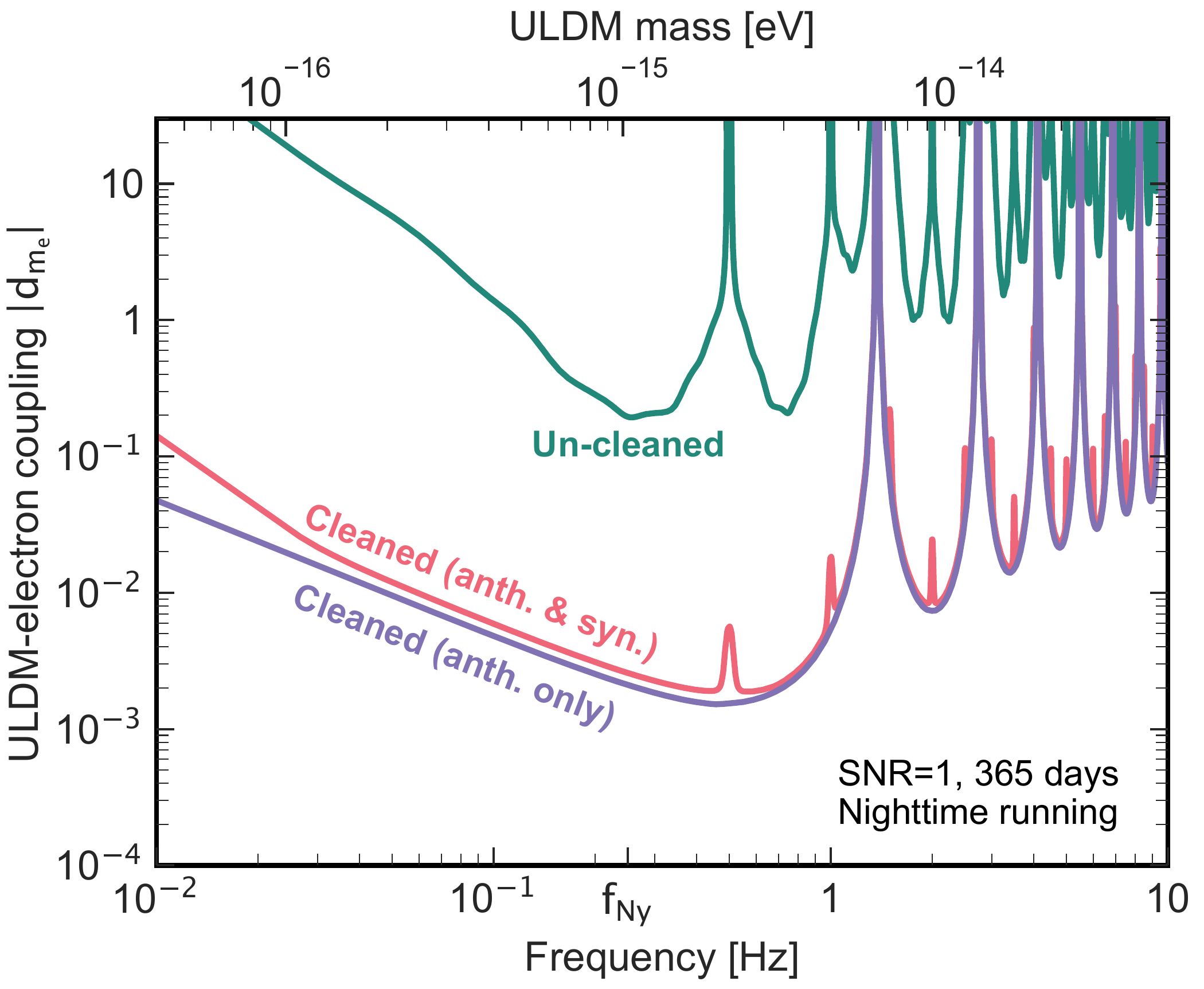}
    \caption{$\mathrm{SNR}=1$ sensitivity projections on the ULDM-electron coupling from simulated data assuming an eight-hour measurement window at nighttime over 365 days.
The green line shows the sensitivity projection from a time series with contributions from anthropogenic and synanthropic noise, along with phase noise from ASN, without any time-series-cleaning strategies applied.
The projections after implementing our cleaning strategy are shown in purple and pink. The purple line includes contributions from anthropogenic noise and phase noise from ASN, while the pink line also includes synanthropic noise.  }
    \label{fig:limits}
\end{figure}

Figure~\ref{fig:limits} shows three sensitivity projections in terms of the ULDM-electron coupling as a function of the ULDM frequency and mass. 
We use the signal-to-noise ratio (SNR) as an estimator of the signal strength and show projections with $\mathrm{SNR}=1$~\cite{Badurina:2021lwr}.
For all three projections, we assume that data is collected during an eight-hour window at nighttime for 365 days.
The green line shows the sensitivity that would be achieved from the raw time series containing ASN with phase noise $\delta \phi =10^{-4}~\mathrm{rad}/\sqrt{\mathrm{Hz}}$ together with anthropogenic and synanthropic noise in the absence of any cleaning strategies.
In contrast, the purple and pink lines demonstrate the sensitivity after masking the transients and detrending the time series.
The purple line assumes the presence of anthropogenic noise, while the pink line assumes that synanthropic noise is also present. 

There are a number of features to note in Fig.~\ref{fig:limits}. Firstly, the impact of the cleaning strategy on the sensitivity projections is dramatic. 
Without cleaning, the sensitivity projection on $d_{m_e}$ is larger by approximately two-to-three orders of magnitude. 
However, with a cleaning strategy in place, the impact of anthropogenic and synanthropic noise can be substantially mitigated.

Secondly, the presence of synanthropic noise leads to a slight reduction in overall sensitivity compared to the anthropogenic-only case.
As discussed in section~\ref{sec:freqenvnoise}, the additional masks applied in the case of synanthropic noise result in a slightly smaller ULDM spike amplitude. 
In the regime where the integration time is smaller than the ULDM coherence time, approximately \SI{0.5}{\hertz} for the chosen parameters, the sensitivity scales with $(T_{\rm{int}})^{-1/2}$. 
Consequently, masking about $45\%$ of the time series instead of $16\%$ reduces the sensitivity by a factor of 1.2, which is consistent with the behaviour observed between approximately \SI{0.04}{\hertz} and \SI{0.5}{\hertz}.
In contrast, above \SI{0.5}{Hz}, the integration time exceeds the ULDM coherence time, causing the sensitivity to scale with $(T_{\rm{int}})^{-1/4}$. As a result, the sensitivity is only reduced by a factor of 1.1, which again corresponds with the observed behaviour in Fig.~\ref{fig:limits}.

Thirdly, we observe a reduction in the sensitivity projection below approximately $0.04\,\mathrm{Hz}$ when synanthropic noise is included, relative to the anthropogenic-only case. 
This is simply a manifestation of the rise in the noise PSD below $0.04\,\mathrm{Hz}$ that we saw in the lower panel of Fig.~\ref{fig:unCutPSD}.
The low-frequency rise in the noise PSD is also responsible for the appearance of Gaussian-like bumps in the pink line at multiples of $2 f_\mathrm{Ny} = \SI{0.5}{\hertz}$. 
Owing to aliasing, the noise PSD below~$f_\mathrm{Ny}$ is folded (or reflected) around~$f_\mathrm{Ny}$. 
Consequently, the rise at low frequencies reappears just below $2 f_\mathrm{Ny}$, and this pattern repeats at higher frequencies. 
A similar behaviour is also observed in the uncleaned line. 
However, in the uncleaned case, the rise at low frequencies is much more pronounced, leading to more prominent bumps at multiples of $2 f_\mathrm{Ny}$.
In contrast, bumps at $2 f_\mathrm{Ny}$ are absent in the anthropogenic-only case (and also in shot-noise limited projections) since the noise remains flat at low frequencies. As a result, when folded, the noise PSD remains flat, implying the absence of any bump features in addition to the expected loss in sensitivity due to the trigonometric functions in eq.~\eqref{eq:DM}.

\begin{figure}
    \centering
    \includegraphics[width=0.93 \columnwidth]{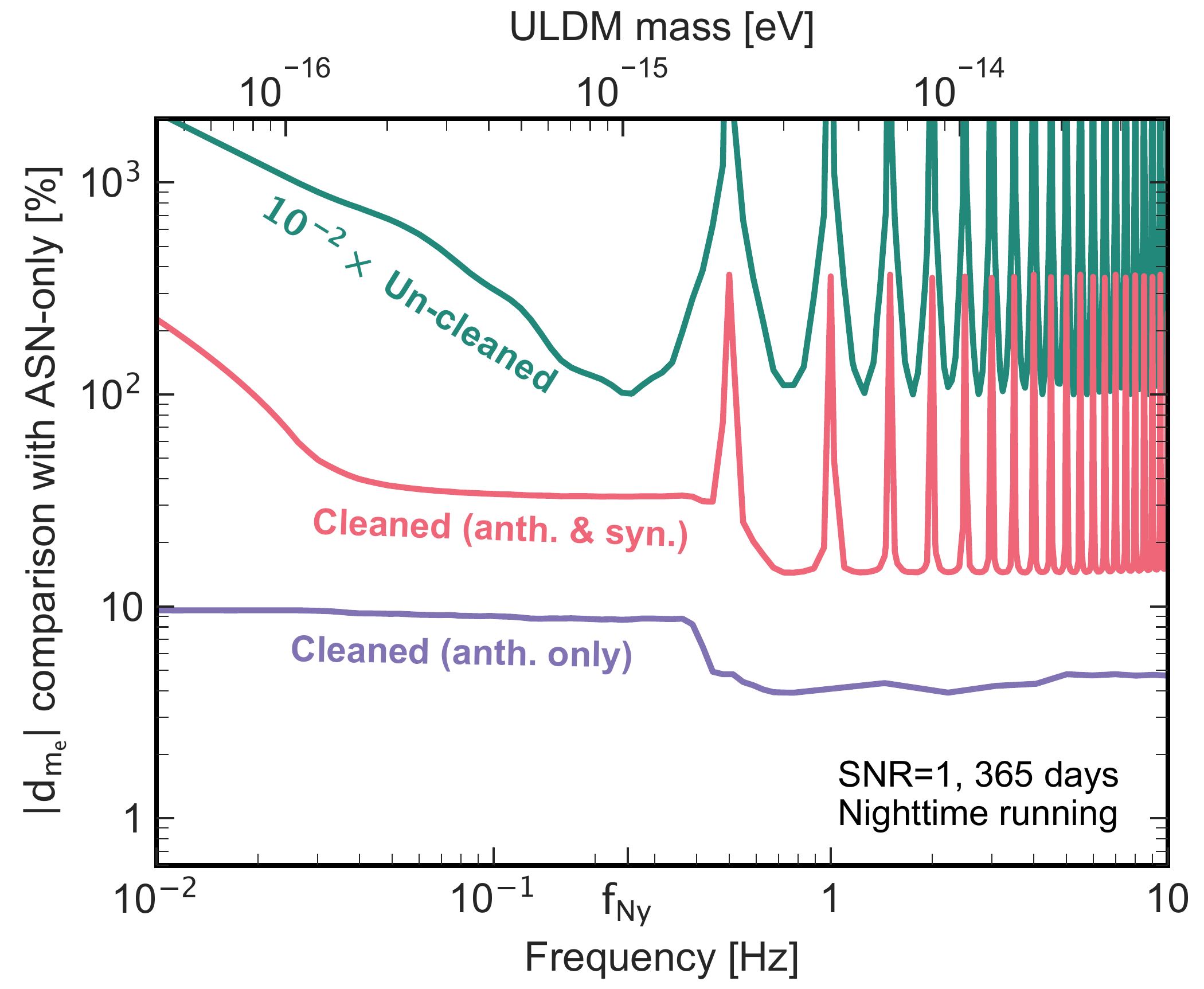}
    \caption{ Relative comparison of $\mathrm{SNR}=1$ sensitivity projections on the ULDM-electron coupling relative to an ASN-limited experiment.
All scenarios assume simulated data from an eight-hour measurement window at nighttime over 365 days.
The green line shows the relative sensitivity for a time series with contributions from anthropogenic and synanthropic noise, along with phase noise from ASN, without any time-series-cleaning strategies applied.
The relative sensitivity after implementing our cleaning strategy is shown in purple and pink for scenarios with anthropogenic-only and anthropogenic and synanthropic noise, respectively.}
    \label{fig:ASNcomp}
\end{figure}

In Fig.~\ref{fig:ASNcomp}, we present a comparison of the $\mathrm{SNR}=1$ sensitivity projections relative to an ASN-limited experiment that collects data during an eight-hour window at nighttime for 365 days. 
The purple line, which represents the anthropogenic only scenario after masking transients and detrending the time series, shows that data cleaning has restored the sensitivity to within 10\% or better of an ASN-limited experiment. The transition at approximately \SI{0.5}{\hertz} occurs because of the different scaling of the sensitivity with~$T_{\rm{int}}$, depending on whether the integration time is greater or smaller than the ULDM coherence time.

The inclusion of synanthropic noise, as shown by the pink line in Fig.~\ref{fig:ASNcomp}, leads to a further reduction in sensitivity since more of the time series is masked. In this scenario, the sensitivity is within 40\% of an ASN-limited experiment over much of the parameter space. Exceptions occur at low frequency and at multiples of $2 f_\mathrm{Ny}$ due to the phenomenon of folding. 
 
Finally, the sensitivity projection obtained using the uncleaned time series, indicated by the green line in Fig.~\ref{fig:ASNcomp}, is substantially weaker. This highlights the importance of noise reduction methods to recover sensitivity to ultra-light dark matter candidates.
 
\section{Summary and outlook \label{sec:summary}}

Forthcoming terrestrial atom interferometer experiments will offer a new probe of dark matter interactions with normal matter.
Among these experiments is AION-10, which will be hosted in the Beecroft Building at the University of Oxford’s Department of Physics.
AION-10 will operate amongst a background of anthropogenic (human sourced) and synanthropic (animal sourced) sources (cf.\ Fig.~\ref{fig:beecroft}).
These sources could induce time-dependent phase shifts that have the potential to hinder searches for the periodic signature produced by ultra-light dark matter.

In this work, we have characterised potential anthropogenic and synanthropic noise sources in the context of AION-10.
Among these sources, we found that massive bodies moving close to AION-10, such as people walking on stairs near the AION-10 supporting tower, can lead to the most significant transients in the phase shift time series (cf.\ Fig.~\ref{fig:PS}).
A passenger lift within the building or vehicles passing on a nearby road would also produce significant transient effects.
While it may be straightforward to implement a detection system to trace the movement of people or vehicles close to the building,
directly monitoring the movement of rodents or other small animals may prove to be more challenging.
To that end, we also characterised the impact of synanthropic sources, which we referred to as random animal transients (RATs) to highlight their potentially untraceable nature.

Searching for ultra-light dark matter with AION-10 will necessitate several months of continuous data taking to enable detection of a dark matter signal with a frequency-domain analysis.
We therefore simulated extended runs of the experiment for periods during the day and night (cf.\ Fig.~\ref{fig:totPS}).
While nighttime running significantly reduced the number of transient events relative to daytime running, additional cleaning of the time series was still required to optimise sensitivity to ultra-light dark matter.

For that reason, we presented a high-level discussion of a data cleaning strategy to mitigate anthropogenic and synanthropic noise sources (cf.\ Fig.~\ref{fig:mask}).
Our strategy involved identifying and masking transient effects, followed by detrending to eliminate any long-term drifts and to establish a common baseline.
While we anticipate that the specific implementation of our cleaning strategy may require modification when applied to real data rather than simulated data, the basic underlying principles should remain the same.

We then performed a frequency-domain analysis on the cleaned time series with the Lomb-Scargle periodogram,
which provides an estimator of the power spectral density for non-uniformly sampled time series.
The cleaned power spectral densities closely resembled the atom shot noise-only case (cf.\ Fig.~\ref{fig:DM_SN} and Fig.~\ref{fig:unCutPSD}),
with only a small increase at low frequency ($\lesssim 0.04\,\mathrm{Hz}$) in the scenario where both anthropogenic and synanthropic noise was present.
This indicated that the data cleaning method successfully identified and removed anthropogenic- and synanthropic-induced transients.

Lastly, we assessed the impact of anthropogenic and synanthropic noise sources on ULDM sensitivity projections (cf.\ Fig.~\ref{fig:limits}). 
In the absence of any data cleaning, the sensitivity deteriorates by two-to-three orders of magnitude compared to an atom shot noise-limited experiment. 
However, through the application of masking transients and detrending the time series, we successfully restored the sensitivity to within 10\% (or 40\%) of an atom shot noise-limited experiment
in the scenario where anthropogenic (or anthropogenic and synanthropic) noise is present (cf.\ Fig.~\ref{fig:ASNcomp}). 

While our findings indicate that masking segments of the time series results in a relatively modest reduction in AION-10's sensitivity to ultra-light dark matter, this approach does come with certain drawbacks. 
Notably, AION-10 is anticipated to operate concurrently with other atom interferometer experiments, like MAGIS-100. 
The utilisation of time-series masks could potentially complicate the process of networking these experiments effectively.

To address these limitations, more sophisticated alternatives may be explored, including the application of machine learning techniques. 
Such techniques could potentially offer enhanced methods of cleaning the time series without relying on masks. 
For instance, a repository of anthropogenic and synanthropic noise templates could be created based on the simulations we have developed in the paper.
Through further feature engineering, these templates could be further refined to train a neural network to accurately extract the noise signal from the time series.
This would preserve the underlying atom shot noise and ultra-light dark matter components intact without losing integration time. 
However, these additional investigations fall beyond the scope of our current study and warrant future research endeavours. 

\section*{Acknowledgements}
We are grateful to members of the AION and MAGIS-100 collaborations for many fruitful discussions, and to Leonardo Badurina, Ankit Beniwal, John Ellis, Jeremiah Mitchell and Tim Kovachy for comments on the manuscript. 
J.C.\ acknowledges support from a King's College London NMES Faculty Studentship. C.M.\ is supported by the Science and Technology Facilities Council (STFC) Grant No.\ ST/T00679X/1. 
For the purpose of open access, the authors have applied a Creative Commons Attribution (CC BY) licence to any Author Accepted Manuscript version arising from this submission. 
No experimental datasets were generated by this research.

\bibliography{main.bbl}

\begin{thebibliography}{104}%
\makeatletter
\providecommand \@ifxundefined [1]{%
 \@ifx{#1\undefined}
}%
\providecommand \@ifnum [1]{%
 \ifnum #1\expandafter \@firstoftwo
 \else \expandafter \@secondoftwo
 \fi
}%
\providecommand \@ifx [1]{%
 \ifx #1\expandafter \@firstoftwo
 \else \expandafter \@secondoftwo
 \fi
}%
\providecommand \natexlab [1]{#1}%
\providecommand \enquote  [1]{``#1''}%
\providecommand \bibnamefont  [1]{#1}%
\providecommand \bibfnamefont [1]{#1}%
\providecommand \citenamefont [1]{#1}%
\providecommand \href@noop [0]{\@secondoftwo}%
\providecommand \href [0]{\begingroup \@sanitize@url \@href}%
\providecommand \@href[1]{\@@startlink{#1}\@@href}%
\providecommand \@@href[1]{\endgroup#1\@@endlink}%
\providecommand \@sanitize@url [0]{\catcode `\\12\catcode `\$12\catcode
  `\&12\catcode `\#12\catcode `\^12\catcode `\_12\catcode `\%12\relax}%
\providecommand \@@startlink[1]{}%
\providecommand \@@endlink[0]{}%
\providecommand \url  [0]{\begingroup\@sanitize@url \@url }%
\providecommand \@url [1]{\endgroup\@href {#1}{\urlprefix }}%
\providecommand \urlprefix  [0]{URL }%
\providecommand \Eprint [0]{\href }%
\providecommand \doibase [0]{https://doi.org/}%
\providecommand \selectlanguage [0]{\@gobble}%
\providecommand \bibinfo  [0]{\@secondoftwo}%
\providecommand \bibfield  [0]{\@secondoftwo}%
\providecommand \translation [1]{[#1]}%
\providecommand \BibitemOpen [0]{}%
\providecommand \bibitemStop [0]{}%
\providecommand \bibitemNoStop [0]{.\EOS\space}%
\providecommand \EOS [0]{\spacefactor3000\relax}%
\providecommand \BibitemShut  [1]{\csname bibitem#1\endcsname}%
\let\auto@bib@innerbib\@empty
\bibitem [{\citenamefont {Buchmueller}\ \emph {et~al.}(2022)\citenamefont
  {Buchmueller} \emph {et~al.}}]{Buchmueller:2022djy}%
  \BibitemOpen
  \bibfield  {author} {\bibinfo {author} {\bibfnamefont {O.}~\bibnamefont
  {Buchmueller}} \emph {et~al.},\ }\bibfield  {title} {\bibinfo {title}
  {{Snowmass 2021: Quantum Sensors for HEP Science -- Interferometers,
  Mechanics, Traps, and Clocks}},\ }in\ \href@noop {} {\emph {\bibinfo
  {booktitle} {{2022 Snowmass Summer Study}}}}\ (\bibinfo {year} {2022})\
  \Eprint {https://arxiv.org/abs/2203.07250} {arXiv:2203.07250 [quant-ph]}
  \BibitemShut {NoStop}%
\bibitem [{\citenamefont {Morel}\ \emph {et~al.}(2020)\citenamefont {Morel},
  \citenamefont {Yao} \emph {et~al.}}]{Morel}%
  \BibitemOpen
  \bibfield  {author} {\bibinfo {author} {\bibfnamefont {L.}~\bibnamefont
  {Morel}}, \bibinfo {author} {\bibfnamefont {Z.}~\bibnamefont {Yao}}, \emph
  {et~al.},\ }\bibfield  {title} {\bibinfo {title} {{Determination of the
  fine-structure constant with an accuracy of 81 parts per trillion}},\ }\href
  {https://doi.org/10.1038/s41586-020-2964-7} {\bibfield  {journal} {\bibinfo
  {journal} {Nature}\ }\textbf {\bibinfo {volume} {588}},\ \bibinfo {pages}
  {61} (\bibinfo {year} {2020})}\BibitemShut {NoStop}%
\bibitem [{\citenamefont {Estey}\ \emph {et~al.}(2015)\citenamefont {Estey},
  \citenamefont {Yu} \emph {et~al.}}]{Estey:2014zha}%
  \BibitemOpen
  \bibfield  {author} {\bibinfo {author} {\bibfnamefont {B.}~\bibnamefont
  {Estey}}, \bibinfo {author} {\bibfnamefont {C.}~\bibnamefont {Yu}}, \emph
  {et~al.},\ }\bibfield  {title} {\bibinfo {title} {{High-Resolution Atom
  Interferometers with Suppressed Diffraction Phases}},\ }\href
  {https://doi.org/10.1103/PhysRevLett.115.083002} {\bibfield  {journal}
  {\bibinfo  {journal} {Phys. Rev. Lett.}\ }\textbf {\bibinfo {volume} {115}},\
  \bibinfo {pages} {083002} (\bibinfo {year} {2015})},\ \Eprint
  {https://arxiv.org/abs/1410.8486} {arXiv:1410.8486 [physics.atom-ph]}
  \BibitemShut {NoStop}%
\bibitem [{\citenamefont {Parker}\ \emph {et~al.}(2018)\citenamefont {Parker},
  \citenamefont {Yu} \emph {et~al.}}]{Parker_2018}%
  \BibitemOpen
  \bibfield  {author} {\bibinfo {author} {\bibfnamefont {R.~H.}\ \bibnamefont
  {Parker}}, \bibinfo {author} {\bibfnamefont {C.}~\bibnamefont {Yu}}, \emph
  {et~al.},\ }\bibfield  {title} {\bibinfo {title} {Measurement of the
  fine-structure constant as a test of the standard model},\ }\href
  {https://doi.org/10.1126/science.aap7706} {\bibfield  {journal} {\bibinfo
  {journal} {Science}\ }\textbf {\bibinfo {volume} {360}},\ \bibinfo {pages}
  {191} (\bibinfo {year} {2018})}\BibitemShut {NoStop}%
\bibitem [{\citenamefont {Arndt}\ and\ \citenamefont
  {Hornberger}(2014)}]{Arndt_2014}%
  \BibitemOpen
  \bibfield  {author} {\bibinfo {author} {\bibfnamefont {M.}~\bibnamefont
  {Arndt}}\ and\ \bibinfo {author} {\bibfnamefont {K.}~\bibnamefont
  {Hornberger}},\ }\bibfield  {title} {\bibinfo {title} {Testing the limits of
  quantum mechanical superpositions},\ }\href
  {https://doi.org/10.1038/nphys2863} {\bibfield  {journal} {\bibinfo
  {journal} {Nature Physics}\ }\textbf {\bibinfo {volume} {10}},\ \bibinfo
  {pages} {271} (\bibinfo {year} {2014})}\BibitemShut {NoStop}%
\bibitem [{\citenamefont {Bassi}\ \emph {et~al.}(2013)\citenamefont {Bassi},
  \citenamefont {Lochan} \emph {et~al.}}]{Bassi:2012bg}%
  \BibitemOpen
  \bibfield  {author} {\bibinfo {author} {\bibfnamefont {A.}~\bibnamefont
  {Bassi}}, \bibinfo {author} {\bibfnamefont {K.}~\bibnamefont {Lochan}}, \emph
  {et~al.},\ }\bibfield  {title} {\bibinfo {title} {{Models of Wave-function
  Collapse, Underlying Theories, and Experimental Tests}},\ }\href
  {https://doi.org/10.1103/RevModPhys.85.471} {\bibfield  {journal} {\bibinfo
  {journal} {Rev. Mod. Phys.}\ }\textbf {\bibinfo {volume} {85}},\ \bibinfo
  {pages} {471} (\bibinfo {year} {2013})},\ \Eprint
  {https://arxiv.org/abs/1204.4325} {arXiv:1204.4325 [quant-ph]} \BibitemShut
  {NoStop}%
\bibitem [{\citenamefont {Manning}\ \emph {et~al.}(2015)\citenamefont
  {Manning}, \citenamefont {Khakimov} \emph {et~al.}}]{Manning:2015cta}%
  \BibitemOpen
  \bibfield  {author} {\bibinfo {author} {\bibfnamefont {A.~G.}\ \bibnamefont
  {Manning}}, \bibinfo {author} {\bibfnamefont {R.~I.}\ \bibnamefont
  {Khakimov}}, \emph {et~al.},\ }\bibfield  {title} {\bibinfo {title}
  {{Wheeler's delayed-choice gedanken experiment with a single atom}},\ }\href
  {https://doi.org/10.1038/nphys3343} {\bibfield  {journal} {\bibinfo
  {journal} {Nature Phys.}\ }\textbf {\bibinfo {volume} {11}},\ \bibinfo
  {pages} {539} (\bibinfo {year} {2015})}\BibitemShut {NoStop}%
\bibitem [{\citenamefont {Roura}(2020)}]{Roura:2018cfg}%
  \BibitemOpen
  \bibfield  {author} {\bibinfo {author} {\bibfnamefont {A.}~\bibnamefont
  {Roura}},\ }\bibfield  {title} {\bibinfo {title} {{Gravitational redshift in
  quantum-clock interferometry}},\ }\href
  {https://doi.org/10.1103/PhysRevX.10.021014} {\bibfield  {journal} {\bibinfo
  {journal} {Phys. Rev. X}\ }\textbf {\bibinfo {volume} {10}},\ \bibinfo
  {pages} {021014} (\bibinfo {year} {2020})},\ \Eprint
  {https://arxiv.org/abs/1810.06744} {arXiv:1810.06744 [physics.atom-ph]}
  \BibitemShut {NoStop}%
\bibitem [{\citenamefont {Zych}\ \emph {et~al.}(2011)\citenamefont {Zych},
  \citenamefont {Costa} \emph {et~al.}}]{Zych:2011hu}%
  \BibitemOpen
  \bibfield  {author} {\bibinfo {author} {\bibfnamefont {M.}~\bibnamefont
  {Zych}}, \bibinfo {author} {\bibfnamefont {F.}~\bibnamefont {Costa}}, \emph
  {et~al.},\ }\bibfield  {title} {\bibinfo {title} {{Quantum interferometric
  visibility as a witness of general relativistic proper time}},\ }\href
  {https://doi.org/10.1038/ncomms1498} {\bibfield  {journal} {\bibinfo
  {journal} {Nature Commun.}\ }\textbf {\bibinfo {volume} {2}},\ \bibinfo
  {pages} {505} (\bibinfo {year} {2011})},\ \Eprint
  {https://arxiv.org/abs/1105.4531} {arXiv:1105.4531 [quant-ph]} \BibitemShut
  {NoStop}%
\bibitem [{\citenamefont {Xu}\ \emph {et~al.}(2019)\citenamefont {Xu},
  \citenamefont {Jaffe} \emph {et~al.}}]{Xu:2019vlt}%
  \BibitemOpen
  \bibfield  {author} {\bibinfo {author} {\bibfnamefont {V.}~\bibnamefont
  {Xu}}, \bibinfo {author} {\bibfnamefont {M.}~\bibnamefont {Jaffe}}, \emph
  {et~al.},\ }\bibfield  {title} {\bibinfo {title} {Probing gravity by holding
  atoms for 20 seconds},\ }\href {https://doi.org/10.1126/science.aay6428}
  {\bibfield  {journal} {\bibinfo  {journal} {Science}\ }\textbf {\bibinfo
  {volume} {366}},\ \bibinfo {pages} {745} (\bibinfo {year}
  {2019})}\BibitemShut {NoStop}%
\bibitem [{\citenamefont {Asenbaum}\ \emph {et~al.}(2020)\citenamefont
  {Asenbaum}, \citenamefont {Overstreet} \emph {et~al.}}]{Asenbaum_2020}%
  \BibitemOpen
  \bibfield  {author} {\bibinfo {author} {\bibfnamefont {P.}~\bibnamefont
  {Asenbaum}}, \bibinfo {author} {\bibfnamefont {C.}~\bibnamefont
  {Overstreet}}, \emph {et~al.},\ }\bibfield  {title} {\bibinfo {title}
  {Atom-interferometric test of the equivalence principle at the $10^{-12}$
  level},\ }\bibfield  {journal} {\bibinfo  {journal} {Physical Review
  Letters}\ }\textbf {\bibinfo {volume} {125}},\ \href
  {https://doi.org/10.1103/physrevlett.125.191101}
  {10.1103/physrevlett.125.191101} (\bibinfo {year} {2020})\BibitemShut
  {NoStop}%
\bibitem [{\citenamefont {Burrage}\ \emph {et~al.}(2015)\citenamefont
  {Burrage}, \citenamefont {Copeland},\ and\ \citenamefont
  {Hinds}}]{Burrage:2014oza}%
  \BibitemOpen
  \bibfield  {author} {\bibinfo {author} {\bibfnamefont {C.}~\bibnamefont
  {Burrage}}, \bibinfo {author} {\bibfnamefont {E.~J.}\ \bibnamefont
  {Copeland}},\ and\ \bibinfo {author} {\bibfnamefont {E.~A.}\ \bibnamefont
  {Hinds}},\ }\bibfield  {title} {\bibinfo {title} {{Probing Dark Energy with
  Atom Interferometry}},\ }\href
  {https://doi.org/10.1088/1475-7516/2015/03/042} {\bibfield  {journal}
  {\bibinfo  {journal} {JCAP}\ }\textbf {\bibinfo {volume} {2015}}\bibfield
  {number} {\bibinfo  {number} { (03)}},\ }\Eprint
  {https://arxiv.org/abs/1408.1409} {arXiv:1408.1409 [astro-ph.CO]}
  \BibitemShut {NoStop}%
\bibitem [{\citenamefont {Hamilton}\ \emph {et~al.}(2015)\citenamefont
  {Hamilton}, \citenamefont {Jaffe} \emph {et~al.}}]{Hamilton:2015zga}%
  \BibitemOpen
  \bibfield  {author} {\bibinfo {author} {\bibfnamefont {P.}~\bibnamefont
  {Hamilton}}, \bibinfo {author} {\bibfnamefont {M.}~\bibnamefont {Jaffe}},
  \emph {et~al.},\ }\bibfield  {title} {\bibinfo {title} {{Atom-interferometry
  constraints on dark energy}},\ }\href
  {https://doi.org/10.1126/science.aaa8883} {\bibfield  {journal} {\bibinfo
  {journal} {Science}\ }\textbf {\bibinfo {volume} {349}},\ \bibinfo {pages}
  {849} (\bibinfo {year} {2015})},\ \Eprint {https://arxiv.org/abs/1502.03888}
  {arXiv:1502.03888 [physics.atom-ph]} \BibitemShut {NoStop}%
\bibitem [{\citenamefont {Elder}\ \emph {et~al.}(2016)\citenamefont {Elder},
  \citenamefont {Khoury} \emph {et~al.}}]{Elder:2016yxm}%
  \BibitemOpen
  \bibfield  {author} {\bibinfo {author} {\bibfnamefont {B.}~\bibnamefont
  {Elder}}, \bibinfo {author} {\bibfnamefont {J.}~\bibnamefont {Khoury}}, \emph
  {et~al.},\ }\bibfield  {title} {\bibinfo {title} {{Chameleon Dark Energy and
  Atom Interferometry}},\ }\href {https://doi.org/10.1103/PhysRevD.94.044051}
  {\bibfield  {journal} {\bibinfo  {journal} {Phys. Rev. D}\ }\textbf {\bibinfo
  {volume} {94}},\ \bibinfo {pages} {044051} (\bibinfo {year} {2016})},\
  \Eprint {https://arxiv.org/abs/1603.06587} {arXiv:1603.06587 [astro-ph.CO]}
  \BibitemShut {NoStop}%
\bibitem [{\citenamefont {Sabulsky}\ \emph {et~al.}(2019)\citenamefont
  {Sabulsky}, \citenamefont {Dutta}, \citenamefont {Hinds}, \citenamefont
  {Elder}, \citenamefont {Burrage},\ and\ \citenamefont
  {Copeland}}]{Sabulsky:2018jma}%
  \BibitemOpen
  \bibfield  {author} {\bibinfo {author} {\bibfnamefont {D.~O.}\ \bibnamefont
  {Sabulsky}}, \bibinfo {author} {\bibfnamefont {I.}~\bibnamefont {Dutta}},
  \bibinfo {author} {\bibfnamefont {E.~A.}\ \bibnamefont {Hinds}}, \bibinfo
  {author} {\bibfnamefont {B.}~\bibnamefont {Elder}}, \bibinfo {author}
  {\bibfnamefont {C.}~\bibnamefont {Burrage}},\ and\ \bibinfo {author}
  {\bibfnamefont {E.~J.}\ \bibnamefont {Copeland}},\ }\bibfield  {title}
  {\bibinfo {title} {{Experiment to detect dark energy forces using atom
  interferometry}},\ }\href {https://doi.org/10.1103/PhysRevLett.123.061102}
  {\bibfield  {journal} {\bibinfo  {journal} {Phys. Rev. Lett.}\ }\textbf
  {\bibinfo {volume} {123}},\ \bibinfo {pages} {061102} (\bibinfo {year}
  {2019})},\ \Eprint {https://arxiv.org/abs/1812.08244} {arXiv:1812.08244
  [physics.atom-ph]} \BibitemShut {NoStop}%
\bibitem [{\citenamefont {Biedermann}\ \emph {et~al.}(2015)\citenamefont
  {Biedermann}, \citenamefont {Wu} \emph {et~al.}}]{Biedermann_2015}%
  \BibitemOpen
  \bibfield  {author} {\bibinfo {author} {\bibfnamefont {G.~W.}\ \bibnamefont
  {Biedermann}}, \bibinfo {author} {\bibfnamefont {X.}~\bibnamefont {Wu}},
  \emph {et~al.},\ }\bibfield  {title} {\bibinfo {title} {Testing gravity with
  cold-atom interferometers},\ }\href
  {https://doi.org/10.1103/PhysRevA.91.033629} {\bibfield  {journal} {\bibinfo
  {journal} {Phys. Rev. A}\ }\textbf {\bibinfo {volume} {91}},\ \bibinfo
  {pages} {033629} (\bibinfo {year} {2015})}\BibitemShut {NoStop}%
\bibitem [{\citenamefont {Rosi}\ \emph {et~al.}(2017)\citenamefont {Rosi},
  \citenamefont {D'Amico} \emph {et~al.}}]{Rosi:2017ieh}%
  \BibitemOpen
  \bibfield  {author} {\bibinfo {author} {\bibfnamefont {G.}~\bibnamefont
  {Rosi}}, \bibinfo {author} {\bibfnamefont {G.}~\bibnamefont {D'Amico}}, \emph
  {et~al.},\ }\bibfield  {title} {\bibinfo {title} {{Quantum test of the
  equivalence principle for atoms in superpositions of internal energy
  eigenstates}},\ }\href {https://doi.org/10.1038/ncomms15529} {\bibfield
  {journal} {\bibinfo  {journal} {Nature Commun.}\ }\textbf {\bibinfo {volume}
  {8}},\ \bibinfo {pages} {5529} (\bibinfo {year} {2017})},\ \Eprint
  {https://arxiv.org/abs/1704.02296} {arXiv:1704.02296 [physics.atom-ph]}
  \BibitemShut {NoStop}%
\bibitem [{\citenamefont {Graham}\ \emph {et~al.}(2016)\citenamefont {Graham},
  \citenamefont {Kaplan} \emph {et~al.}}]{Graham:2015ifn}%
  \BibitemOpen
  \bibfield  {author} {\bibinfo {author} {\bibfnamefont {P.~W.}\ \bibnamefont
  {Graham}}, \bibinfo {author} {\bibfnamefont {D.~E.}\ \bibnamefont {Kaplan}},
  \emph {et~al.},\ }\bibfield  {title} {\bibinfo {title} {{Dark Matter Direct
  Detection with Accelerometers}},\ }\href
  {https://doi.org/10.1103/PhysRevD.93.075029} {\bibfield  {journal} {\bibinfo
  {journal} {Phys. Rev. D}\ }\textbf {\bibinfo {volume} {93}},\ \bibinfo
  {pages} {075029} (\bibinfo {year} {2016})},\ \Eprint
  {https://arxiv.org/abs/1512.06165} {arXiv:1512.06165 [hep-ph]} \BibitemShut
  {NoStop}%
\bibitem [{\citenamefont {Geraci}\ and\ \citenamefont
  {Derevianko}(2016)}]{Geraci:2016fva}%
  \BibitemOpen
  \bibfield  {author} {\bibinfo {author} {\bibfnamefont {A.~A.}\ \bibnamefont
  {Geraci}}\ and\ \bibinfo {author} {\bibfnamefont {A.}~\bibnamefont
  {Derevianko}},\ }\bibfield  {title} {\bibinfo {title} {{Sensitivity of atom
  interferometry to ultralight scalar field dark matter}},\ }\href
  {https://doi.org/10.1103/PhysRevLett.117.261301} {\bibfield  {journal}
  {\bibinfo  {journal} {Phys. Rev. Lett.}\ }\textbf {\bibinfo {volume} {117}},\
  \bibinfo {pages} {261301} (\bibinfo {year} {2016})},\ \Eprint
  {https://arxiv.org/abs/1605.04048} {arXiv:1605.04048 [physics.atom-ph]}
  \BibitemShut {NoStop}%
\bibitem [{\citenamefont {Arvanitaki}\ \emph {et~al.}(2018)\citenamefont
  {Arvanitaki}, \citenamefont {Graham} \emph {et~al.}}]{Arvanitaki:2016fyj}%
  \BibitemOpen
  \bibfield  {author} {\bibinfo {author} {\bibfnamefont {A.}~\bibnamefont
  {Arvanitaki}}, \bibinfo {author} {\bibfnamefont {P.~W.}\ \bibnamefont
  {Graham}}, \emph {et~al.},\ }\bibfield  {title} {\bibinfo {title} {{Search
  for light scalar dark matter with atomic gravitational wave detectors}},\
  }\href {https://doi.org/10.1103/PhysRevD.97.075020} {\bibfield  {journal}
  {\bibinfo  {journal} {Phys. Rev. D}\ }\textbf {\bibinfo {volume} {97}},\
  \bibinfo {pages} {075020} (\bibinfo {year} {2018})},\ \Eprint
  {https://arxiv.org/abs/1606.04541} {arXiv:1606.04541 [hep-ph]} \BibitemShut
  {NoStop}%
\bibitem [{\citenamefont {Badurina}\ \emph {et~al.}(2022)\citenamefont
  {Badurina}, \citenamefont {Blas},\ and\ \citenamefont
  {McCabe}}]{Badurina:2021lwr}%
  \BibitemOpen
  \bibfield  {author} {\bibinfo {author} {\bibfnamefont {L.}~\bibnamefont
  {Badurina}}, \bibinfo {author} {\bibfnamefont {D.}~\bibnamefont {Blas}},\
  and\ \bibinfo {author} {\bibfnamefont {C.}~\bibnamefont {McCabe}},\
  }\bibfield  {title} {\bibinfo {title} {{Refined ultralight scalar dark matter
  searches with compact atom gradiometers}},\ }\href
  {https://doi.org/10.1103/PhysRevD.105.023006} {\bibfield  {journal} {\bibinfo
   {journal} {Phys. Rev. D}\ }\textbf {\bibinfo {volume} {105}},\ \bibinfo
  {pages} {023006} (\bibinfo {year} {2022})},\ \Eprint
  {https://arxiv.org/abs/2109.10965} {arXiv:2109.10965 [astro-ph.CO]}
  \BibitemShut {NoStop}%
\bibitem [{\citenamefont {Badurina}\ \emph {et~al.}(2021)\citenamefont
  {Badurina}, \citenamefont {Buchmueller}, \citenamefont {Ellis}, \citenamefont
  {Lewicki}, \citenamefont {McCabe},\ and\ \citenamefont
  {Vaskonen}}]{Badurina:2021rgt}%
  \BibitemOpen
  \bibfield  {author} {\bibinfo {author} {\bibfnamefont {L.}~\bibnamefont
  {Badurina}}, \bibinfo {author} {\bibfnamefont {O.}~\bibnamefont
  {Buchmueller}}, \bibinfo {author} {\bibfnamefont {J.}~\bibnamefont {Ellis}},
  \bibinfo {author} {\bibfnamefont {M.}~\bibnamefont {Lewicki}}, \bibinfo
  {author} {\bibfnamefont {C.}~\bibnamefont {McCabe}},\ and\ \bibinfo {author}
  {\bibfnamefont {V.}~\bibnamefont {Vaskonen}},\ }\bibfield  {title} {\bibinfo
  {title} {{Prospective sensitivities of atom interferometers to gravitational
  waves and ultralight dark matter}},\ }\href
  {https://doi.org/10.1098/rsta.2021.0060} {\bibfield  {journal} {\bibinfo
  {journal} {Phil. Trans. A. Math. Phys. Eng. Sci.}\ }\textbf {\bibinfo
  {volume} {380}},\ \bibinfo {pages} {20210060} (\bibinfo {year} {2021})},\
  \Eprint {https://arxiv.org/abs/2108.02468} {arXiv:2108.02468 [gr-qc]}
  \BibitemShut {NoStop}%
\bibitem [{\citenamefont {Badurina}\ \emph
  {et~al.}(2023{\natexlab{a}})\citenamefont {Badurina}, \citenamefont {Gibson},
  \citenamefont {McCabe},\ and\ \citenamefont {Mitchell}}]{Badurina:2022ngn}%
  \BibitemOpen
  \bibfield  {author} {\bibinfo {author} {\bibfnamefont {L.}~\bibnamefont
  {Badurina}}, \bibinfo {author} {\bibfnamefont {V.}~\bibnamefont {Gibson}},
  \bibinfo {author} {\bibfnamefont {C.}~\bibnamefont {McCabe}},\ and\ \bibinfo
  {author} {\bibfnamefont {J.}~\bibnamefont {Mitchell}},\ }\bibfield  {title}
  {\bibinfo {title} {{Ultralight dark matter searches at the sub-Hz frontier
  with atom multigradiometry}},\ }\href
  {https://doi.org/10.1103/PhysRevD.107.055002} {\bibfield  {journal} {\bibinfo
   {journal} {Phys. Rev. D}\ }\textbf {\bibinfo {volume} {107}},\ \bibinfo
  {pages} {055002} (\bibinfo {year} {2023}{\natexlab{a}})},\ \Eprint
  {https://arxiv.org/abs/2211.01854} {arXiv:2211.01854 [hep-ph]} \BibitemShut
  {NoStop}%
\bibitem [{\citenamefont {Di~Pumpo}\ \emph {et~al.}(2022)\citenamefont
  {Di~Pumpo}, \citenamefont {Friedrich} \emph {et~al.}}]{DiPumpo:2022muv}%
  \BibitemOpen
  \bibfield  {author} {\bibinfo {author} {\bibfnamefont {F.}~\bibnamefont
  {Di~Pumpo}}, \bibinfo {author} {\bibfnamefont {A.}~\bibnamefont {Friedrich}},
  \emph {et~al.},\ }\bibfield  {title} {\bibinfo {title} {{Light propagation
  and atom interferometry in gravity and dilaton fields}},\ }\href
  {https://doi.org/10.1103/PhysRevD.105.084065} {\bibfield  {journal} {\bibinfo
   {journal} {Phys. Rev. D}\ }\textbf {\bibinfo {volume} {105}},\ \bibinfo
  {pages} {084065} (\bibinfo {year} {2022})},\ \Eprint
  {https://arxiv.org/abs/2201.07053} {arXiv:2201.07053 [quant-ph]} \BibitemShut
  {NoStop}%
\bibitem [{\citenamefont {Dimopoulos}\ \emph {et~al.}(2008)\citenamefont
  {Dimopoulos}, \citenamefont {Graham} \emph {et~al.}}]{Dimopoulos:2008sv}%
  \BibitemOpen
  \bibfield  {author} {\bibinfo {author} {\bibfnamefont {S.}~\bibnamefont
  {Dimopoulos}}, \bibinfo {author} {\bibfnamefont {P.~W.}\ \bibnamefont
  {Graham}}, \emph {et~al.},\ }\bibfield  {title} {\bibinfo {title} {{An Atomic
  Gravitational Wave Interferometric Sensor (AGIS)}},\ }\href
  {https://doi.org/10.1103/PhysRevD.78.122002} {\bibfield  {journal} {\bibinfo
  {journal} {Phys. Rev. D}\ }\textbf {\bibinfo {volume} {78}},\ \bibinfo
  {pages} {122002} (\bibinfo {year} {2008})},\ \Eprint
  {https://arxiv.org/abs/0806.2125} {arXiv:0806.2125 [gr-qc]} \BibitemShut
  {NoStop}%
\bibitem [{\citenamefont {Dimopoulos}\ \emph {et~al.}(2009)\citenamefont
  {Dimopoulos}, \citenamefont {Graham} \emph {et~al.}}]{Dimopoulos:2007cj}%
  \BibitemOpen
  \bibfield  {author} {\bibinfo {author} {\bibfnamefont {S.}~\bibnamefont
  {Dimopoulos}}, \bibinfo {author} {\bibfnamefont {P.~W.}\ \bibnamefont
  {Graham}}, \emph {et~al.},\ }\bibfield  {title} {\bibinfo {title}
  {{Gravitational Wave Detection with Atom Interferometry}},\ }\href
  {https://doi.org/10.1016/j.physletb.2009.06.011} {\bibfield  {journal}
  {\bibinfo  {journal} {Phys. Lett. B}\ }\textbf {\bibinfo {volume} {678}},\
  \bibinfo {pages} {37} (\bibinfo {year} {2009})},\ \Eprint
  {https://arxiv.org/abs/0712.1250} {arXiv:0712.1250 [gr-qc]} \BibitemShut
  {NoStop}%
\bibitem [{\citenamefont {Yu}\ and\ \citenamefont
  {Tinto}(2011{\natexlab{a}})}]{Yu:2010ss}%
  \BibitemOpen
  \bibfield  {author} {\bibinfo {author} {\bibfnamefont {N.}~\bibnamefont
  {Yu}}\ and\ \bibinfo {author} {\bibfnamefont {M.}~\bibnamefont {Tinto}},\
  }\bibfield  {title} {\bibinfo {title} {{Gravitational wave detection with
  single-laser atom interferometers}},\ }\href
  {https://doi.org/10.1007/s10714-010-1055-8} {\bibfield  {journal} {\bibinfo
  {journal} {Gen. Rel. Grav.}\ }\textbf {\bibinfo {volume} {43}},\ \bibinfo
  {pages} {1943} (\bibinfo {year} {2011}{\natexlab{a}})},\ \Eprint
  {https://arxiv.org/abs/1003.4218} {arXiv:1003.4218 [gr-qc]} \BibitemShut
  {NoStop}%
\bibitem [{\citenamefont {Chaibi}\ \emph {et~al.}(2016)\citenamefont {Chaibi},
  \citenamefont {Geiger} \emph {et~al.}}]{Chaibi:2016dze}%
  \BibitemOpen
  \bibfield  {author} {\bibinfo {author} {\bibfnamefont {W.}~\bibnamefont
  {Chaibi}}, \bibinfo {author} {\bibfnamefont {R.}~\bibnamefont {Geiger}},
  \emph {et~al.},\ }\bibfield  {title} {\bibinfo {title} {{Low Frequency
  Gravitational Wave Detection With Ground Based Atom Interferometer Arrays}},\
  }\href {https://doi.org/10.1103/PhysRevD.93.021101} {\bibfield  {journal}
  {\bibinfo  {journal} {Phys. Rev. D}\ }\textbf {\bibinfo {volume} {93}},\
  \bibinfo {pages} {021101} (\bibinfo {year} {2016})},\ \Eprint
  {https://arxiv.org/abs/1601.00417} {arXiv:1601.00417 [physics.atom-ph]}
  \BibitemShut {NoStop}%
\bibitem [{\citenamefont {Graham}\ \emph {et~al.}(2017)\citenamefont {Graham},
  \citenamefont {Hogan} \emph {et~al.}}]{Graham:2017pmn}%
  \BibitemOpen
  \bibfield  {author} {\bibinfo {author} {\bibfnamefont {P.~W.}\ \bibnamefont
  {Graham}}, \bibinfo {author} {\bibfnamefont {J.~M.}\ \bibnamefont {Hogan}},
  \emph {et~al.} (\bibinfo {collaboration} {MAGIS}),\ }\bibfield  {title}
  {\bibinfo {title} {{Mid-band gravitational wave detection with precision
  atomic sensors}}\ }(\bibinfo {year} {2017})\ \Eprint
  {https://arxiv.org/abs/1711.02225} {arXiv:1711.02225 [astro-ph.IM]}
  \BibitemShut {NoStop}%
\bibitem [{\citenamefont {Loriani}\ \emph {et~al.}(2019)\citenamefont {Loriani}
  \emph {et~al.}}]{Loriani:2018qej}%
  \BibitemOpen
  \bibfield  {author} {\bibinfo {author} {\bibfnamefont {S.}~\bibnamefont
  {Loriani}} \emph {et~al.},\ }\bibfield  {title} {\bibinfo {title} {{Atomic
  source selection in space-borne gravitational wave detection}},\ }\href
  {https://doi.org/10.1088/1367-2630/ab22d0} {\bibfield  {journal} {\bibinfo
  {journal} {New J. Phys.}\ }\textbf {\bibinfo {volume} {21}},\ \bibinfo
  {pages} {063030} (\bibinfo {year} {2019})},\ \Eprint
  {https://arxiv.org/abs/1812.11348} {arXiv:1812.11348 [physics.atom-ph]}
  \BibitemShut {NoStop}%
\bibitem [{\citenamefont {Schubert}\ \emph {et~al.}(2019)\citenamefont
  {Schubert}, \citenamefont {Schlippert} \emph {et~al.}}]{Schubert:2019ycf}%
  \BibitemOpen
  \bibfield  {author} {\bibinfo {author} {\bibfnamefont {C.}~\bibnamefont
  {Schubert}}, \bibinfo {author} {\bibfnamefont {D.}~\bibnamefont
  {Schlippert}}, \emph {et~al.},\ }\bibfield  {title} {\bibinfo {title}
  {{Scalable, symmetric atom interferometer for infrasound gravitational wave
  detection}}\ }(\bibinfo {year} {2019})\ \Eprint
  {https://arxiv.org/abs/1909.01951} {arXiv:1909.01951 [quant-ph]} \BibitemShut
  {NoStop}%
\bibitem [{\citenamefont {Canuel}\ \emph {et~al.}(2006)\citenamefont {Canuel},
  \citenamefont {Leduc} \emph {et~al.}}]{Canuel_2006}%
  \BibitemOpen
  \bibfield  {author} {\bibinfo {author} {\bibfnamefont {B.}~\bibnamefont
  {Canuel}}, \bibinfo {author} {\bibfnamefont {F.}~\bibnamefont {Leduc}}, \emph
  {et~al.},\ }\bibfield  {title} {\bibinfo {title} {Six-axis inertial sensor
  using cold-atom interferometry},\ }\bibfield  {journal} {\bibinfo  {journal}
  {Physical Review Letters}\ }\textbf {\bibinfo {volume} {97}},\ \href
  {https://doi.org/10.1103/physrevlett.97.010402}
  {10.1103/physrevlett.97.010402} (\bibinfo {year} {2006})\BibitemShut
  {NoStop}%
\bibitem [{\citenamefont {Hu}\ \emph {et~al.}(2019)\citenamefont {Hu},
  \citenamefont {Wang} \emph {et~al.}}]{Hu_2019}%
  \BibitemOpen
  \bibfield  {author} {\bibinfo {author} {\bibfnamefont {L.}~\bibnamefont
  {Hu}}, \bibinfo {author} {\bibfnamefont {E.}~\bibnamefont {Wang}}, \emph
  {et~al.},\ }\bibfield  {title} {\bibinfo {title} {Sr atom interferometry with
  the optical clock transition as a gravimeter and a gravity gradiometer},\
  }\href {https://doi.org/10.1088/1361-6382/ab4d18} {\bibfield  {journal}
  {\bibinfo  {journal} {Classical and Quantum Gravity}\ }\textbf {\bibinfo
  {volume} {37}},\ \bibinfo {pages} {014001} (\bibinfo {year}
  {2019})}\BibitemShut {NoStop}%
\bibitem [{\citenamefont {Buchmueller}\ \emph {et~al.}(2023)\citenamefont
  {Buchmueller}, \citenamefont {Ellis},\ and\ \citenamefont
  {Schneider}}]{Buchmueller:2023nll}%
  \BibitemOpen
  \bibfield  {author} {\bibinfo {author} {\bibfnamefont {O.}~\bibnamefont
  {Buchmueller}}, \bibinfo {author} {\bibfnamefont {J.}~\bibnamefont {Ellis}},\
  and\ \bibinfo {author} {\bibfnamefont {U.}~\bibnamefont {Schneider}},\
  }\bibfield  {title} {\bibinfo {title} {{Large-Scale Atom Interferometry for
  Fundamental Physics}}\ }(\bibinfo {year} {2023})\ \Eprint
  {https://arxiv.org/abs/2306.17726} {arXiv:2306.17726 [astro-ph.CO]}
  \BibitemShut {NoStop}%
\bibitem [{\citenamefont {Badurina}\ \emph {et~al.}(2020)\citenamefont
  {Badurina} \emph {et~al.}}]{Badurina:2019hst}%
  \BibitemOpen
  \bibfield  {author} {\bibinfo {author} {\bibfnamefont {L.}~\bibnamefont
  {Badurina}} \emph {et~al.},\ }\bibfield  {title} {\bibinfo {title} {{AION: An
  Atom Interferometer Observatory and Network}},\ }\href
  {https://doi.org/10.1088/1475-7516/2020/05/011} {\bibfield  {journal}
  {\bibinfo  {journal} {JCAP}\ }\textbf {\bibinfo {volume} {2020}}\bibfield
  {number} {\bibinfo  {number} { (05)},\ \bibinfo {pages} {011}},\ }\Eprint
  {https://arxiv.org/abs/1911.11755} {arXiv:1911.11755 [astro-ph.CO]}
  \BibitemShut {NoStop}%
\bibitem [{\citenamefont {Canuel}\ \emph {et~al.}(2020)\citenamefont {Canuel},
  \citenamefont {Abend} \emph {et~al.}}]{elgar}%
  \BibitemOpen
  \bibfield  {author} {\bibinfo {author} {\bibfnamefont {B.}~\bibnamefont
  {Canuel}}, \bibinfo {author} {\bibfnamefont {S.}~\bibnamefont {Abend}}, \emph
  {et~al.},\ }\href {https://doi.org/10.48550/ARXIV.2007.04014} {\bibinfo
  {title} {Technologies for the {ELGAR} large scale atom interferometer array}}
  (\bibinfo {year} {2020})\BibitemShut {NoStop}%
\bibitem [{\citenamefont {Abe}\ \emph {et~al.}(2021)\citenamefont {Abe} \emph
  {et~al.}}]{MAGIS-100:2021etm}%
  \BibitemOpen
  \bibfield  {author} {\bibinfo {author} {\bibfnamefont {M.}~\bibnamefont
  {Abe}} \emph {et~al.} (\bibinfo {collaboration} {MAGIS-100}),\ }\bibfield
  {title} {\bibinfo {title} {{Matter-wave Atomic Gradiometer Interferometric
  Sensor (MAGIS-100)}},\ }\href {https://doi.org/10.1088/2058-9565/abf719}
  {\bibfield  {journal} {\bibinfo  {journal} {Quantum Sci. Technol.}\ }\textbf
  {\bibinfo {volume} {6}},\ \bibinfo {pages} {044003} (\bibinfo {year}
  {2021})},\ \Eprint {https://arxiv.org/abs/2104.02835} {arXiv:2104.02835
  [physics.atom-ph]} \BibitemShut {NoStop}%
\bibitem [{\citenamefont {Canuel}\ \emph {et~al.}(2018)\citenamefont {Canuel},
  \citenamefont {Bertoldi} \emph {et~al.}}]{miga}%
  \BibitemOpen
  \bibfield  {author} {\bibinfo {author} {\bibfnamefont {B.}~\bibnamefont
  {Canuel}}, \bibinfo {author} {\bibfnamefont {A.}~\bibnamefont {Bertoldi}},
  \emph {et~al.},\ }\bibfield  {title} {\bibinfo {title} {Exploring gravity
  with the {MIGA} large scale atom interferometer},\ }\bibfield  {journal}
  {\bibinfo  {journal} {Scientific Reports}\ }\textbf {\bibinfo {volume} {8}},\
  \href {https://doi.org/10.1038/s41598-018-32165-z}
  {10.1038/s41598-018-32165-z} (\bibinfo {year} {2018})\BibitemShut {NoStop}%
\bibitem [{\citenamefont {Zhan}\ \emph {et~al.}(2019)\citenamefont {Zhan},
  \citenamefont {Wang} \emph {et~al.}}]{Zaiga}%
  \BibitemOpen
  \bibfield  {author} {\bibinfo {author} {\bibfnamefont {M.-S.}\ \bibnamefont
  {Zhan}}, \bibinfo {author} {\bibfnamefont {J.}~\bibnamefont {Wang}}, \emph
  {et~al.},\ }\bibfield  {title} {\bibinfo {title} {{ZAIGA}: Zhaoshan
  long-baseline atom interferometer gravitation antenna},\ }\href
  {https://doi.org/10.1142/s0218271819400054} {\bibfield  {journal} {\bibinfo
  {journal} {International Journal of Modern Physics D}\ }\textbf {\bibinfo
  {volume} {29}},\ \bibinfo {pages} {1940005} (\bibinfo {year}
  {2019})}\BibitemShut {NoStop}%
\bibitem [{\citenamefont {El-Neaj}\ \emph {et~al.}(2020)\citenamefont
  {El-Neaj}, \citenamefont {Alpigiani} \emph {et~al.}}]{aedge}%
  \BibitemOpen
  \bibfield  {author} {\bibinfo {author} {\bibfnamefont {Y.~A.}\ \bibnamefont
  {El-Neaj}}, \bibinfo {author} {\bibfnamefont {C.}~\bibnamefont {Alpigiani}},
  \emph {et~al.},\ }\bibfield  {title} {\bibinfo {title} {{AEDGE}: Atomic
  experiment for dark matter and gravity exploration in space},\ }\bibfield
  {journal} {\bibinfo  {journal} {{EPJ} Quantum Technology}\ }\textbf {\bibinfo
  {volume} {7}},\ \href {https://doi.org/10.1140/epjqt/s40507-020-0080-0}
  {10.1140/epjqt/s40507-020-0080-0} (\bibinfo {year} {2020})\BibitemShut
  {NoStop}%
\bibitem [{\citenamefont {Aguilera}\ \emph {et~al.}(2014)\citenamefont
  {Aguilera}, \citenamefont {Ahlers} \emph {et~al.}}]{ste-quest}%
  \BibitemOpen
  \bibfield  {author} {\bibinfo {author} {\bibfnamefont {D.~N.}\ \bibnamefont
  {Aguilera}}, \bibinfo {author} {\bibfnamefont {H.}~\bibnamefont {Ahlers}},
  \emph {et~al.},\ }\bibfield  {title} {\bibinfo {title}
  {{STE}-{QUEST}{\textemdash}test of the universality of free fall using cold
  atom interferometry},\ }\href
  {https://doi.org/10.1088/0264-9381/31/11/115010} {\bibfield  {journal}
  {\bibinfo  {journal} {Classical and Quantum Gravity}\ }\textbf {\bibinfo
  {volume} {31}},\ \bibinfo {pages} {115010} (\bibinfo {year}
  {2014})}\BibitemShut {NoStop}%
\bibitem [{\citenamefont {Alonso}\ \emph {et~al.}(2022)\citenamefont {Alonso}
  \emph {et~al.}}]{Alonso:2022oot}%
  \BibitemOpen
  \bibfield  {author} {\bibinfo {author} {\bibfnamefont {I.}~\bibnamefont
  {Alonso}} \emph {et~al.},\ }\bibfield  {title} {\bibinfo {title} {{Cold atoms
  in space: community workshop summary and proposed road-map}},\ }\href
  {https://doi.org/10.1140/epjqt/s40507-022-00147-w} {\bibfield  {journal}
  {\bibinfo  {journal} {EPJ Quant. Technol.}\ }\textbf {\bibinfo {volume}
  {9}},\ \bibinfo {pages} {30} (\bibinfo {year} {2022})},\ \Eprint
  {https://arxiv.org/abs/2201.07789} {arXiv:2201.07789 [astro-ph.IM]}
  \BibitemShut {NoStop}%
\bibitem [{\citenamefont {Peters}\ \emph
  {et~al.}(2001{\natexlab{a}})\citenamefont {Peters}, \citenamefont {Chung},\
  and\ \citenamefont {Chu}}]{Peters_2001}%
  \BibitemOpen
  \bibfield  {author} {\bibinfo {author} {\bibfnamefont {A.}~\bibnamefont
  {Peters}}, \bibinfo {author} {\bibfnamefont {K.}~\bibnamefont {Chung}},\ and\
  \bibinfo {author} {\bibfnamefont {S.}~\bibnamefont {Chu}},\ }\bibfield
  {title} {\bibinfo {title} {High-precision gravity measurements using atom
  interferometry},\ }\href {https://doi.org/10.1088/0026-1394/38/1/4}
  {\bibfield  {journal} {\bibinfo  {journal} {Metrologia}\ }\textbf {\bibinfo
  {volume} {38}},\ \bibinfo {pages} {25} (\bibinfo {year}
  {2001}{\natexlab{a}})}\BibitemShut {NoStop}%
\bibitem [{\citenamefont {Kasevich}\ and\ \citenamefont
  {Chu}(1992)}]{Kasevich_1992}%
  \BibitemOpen
  \bibfield  {author} {\bibinfo {author} {\bibfnamefont {M.}~\bibnamefont
  {Kasevich}}\ and\ \bibinfo {author} {\bibfnamefont {S.}~\bibnamefont {Chu}},\
  }\bibfield  {title} {\bibinfo {title} {Measurement of the gravitational
  acceleration of an atom with a light-pulse atom interferometer},\ }\href
  {https://doi.org/10.1007/BF00325375} {\bibfield  {journal} {\bibinfo
  {journal} {Applied Physics B}\ }\textbf {\bibinfo {volume} {54}},\ \bibinfo
  {pages} {321} (\bibinfo {year} {1992})}\BibitemShut {NoStop}%
\bibitem [{\citenamefont {McGuirk}\ \emph {et~al.}(2002)\citenamefont
  {McGuirk}, \citenamefont {Foster} \emph {et~al.}}]{McGuirk2002}%
  \BibitemOpen
  \bibfield  {author} {\bibinfo {author} {\bibfnamefont {J.~M.}\ \bibnamefont
  {McGuirk}}, \bibinfo {author} {\bibfnamefont {G.~T.}\ \bibnamefont {Foster}},
  \emph {et~al.},\ }\bibfield  {title} {\bibinfo {title} {Sensitive
  absolute-gravity gradiometry using atom interferometry},\ }\href
  {https://doi.org/10.1103/PhysRevA.65.033608} {\bibfield  {journal} {\bibinfo
  {journal} {Phys. Rev. A}\ }\textbf {\bibinfo {volume} {65}},\ \bibinfo
  {pages} {033608} (\bibinfo {year} {2002})}\BibitemShut {NoStop}%
\bibitem [{\citenamefont {Asenbaum}\ \emph {et~al.}(2017)\citenamefont
  {Asenbaum}, \citenamefont {Overstreet} \emph {et~al.}}]{Asenbaum_2017}%
  \BibitemOpen
  \bibfield  {author} {\bibinfo {author} {\bibfnamefont {P.}~\bibnamefont
  {Asenbaum}}, \bibinfo {author} {\bibfnamefont {C.}~\bibnamefont
  {Overstreet}}, \emph {et~al.},\ }\bibfield  {title} {\bibinfo {title} {Phase
  shift in an atom interferometer due to spacetime curvature across its wave
  function},\ }\bibfield  {journal} {\bibinfo  {journal} {Physical Review
  Letters}\ }\textbf {\bibinfo {volume} {118}},\ \href
  {https://doi.org/10.1103/physrevlett.118.183602}
  {10.1103/physrevlett.118.183602} (\bibinfo {year} {2017})\BibitemShut
  {NoStop}%
\bibitem [{\citenamefont {Fixler}\ \emph {et~al.}(2007)\citenamefont {Fixler},
  \citenamefont {Foster} \emph {et~al.}}]{Fixler_2007}%
  \BibitemOpen
  \bibfield  {author} {\bibinfo {author} {\bibfnamefont {J.~B.}\ \bibnamefont
  {Fixler}}, \bibinfo {author} {\bibfnamefont {G.~T.}\ \bibnamefont {Foster}},
  \emph {et~al.},\ }\bibfield  {title} {\bibinfo {title} {Atom interferometer
  measurement of the newtonian constant of gravity},\ }\href
  {https://doi.org/10.1126/science.1135459} {\bibfield  {journal} {\bibinfo
  {journal} {Science}\ }\textbf {\bibinfo {volume} {315}},\ \bibinfo {pages}
  {74} (\bibinfo {year} {2007})}\BibitemShut {NoStop}%
\bibitem [{\citenamefont {Rosi}\ \emph {et~al.}(2014)\citenamefont {Rosi},
  \citenamefont {Sorrentino} \emph {et~al.}}]{Rosi_2014}%
  \BibitemOpen
  \bibfield  {author} {\bibinfo {author} {\bibfnamefont {G.}~\bibnamefont
  {Rosi}}, \bibinfo {author} {\bibfnamefont {F.}~\bibnamefont {Sorrentino}},
  \emph {et~al.},\ }\bibfield  {title} {\bibinfo {title} {Precision measurement
  of the newtonian gravitational constant using cold atoms},\ }\href
  {https://doi.org/10.1038/nature13433} {\bibfield  {journal} {\bibinfo
  {journal} {Nature}\ }\textbf {\bibinfo {volume} {510}},\ \bibinfo {pages}
  {518} (\bibinfo {year} {2014})}\BibitemShut {NoStop}%
\bibitem [{\citenamefont {Stray}\ \emph {et~al.}(2022)\citenamefont {Stray},
  \citenamefont {Lamb} \emph {et~al.}}]{Stray_2022}%
  \BibitemOpen
  \bibfield  {author} {\bibinfo {author} {\bibfnamefont {B.}~\bibnamefont
  {Stray}}, \bibinfo {author} {\bibfnamefont {A.}~\bibnamefont {Lamb}}, \emph
  {et~al.},\ }\bibfield  {title} {\bibinfo {title} {Quantum sensing for gravity
  cartography},\ }\href {https://doi.org/10.1038/s41586-021-04315-3} {\bibfield
   {journal} {\bibinfo  {journal} {Nature}\ }\textbf {\bibinfo {volume}
  {602}},\ \bibinfo {pages} {590} (\bibinfo {year} {2022})}\BibitemShut
  {NoStop}%
\bibitem [{\citenamefont {Creighton}(2008)}]{Creighton_2008}%
  \BibitemOpen
  \bibfield  {author} {\bibinfo {author} {\bibfnamefont {T.}~\bibnamefont
  {Creighton}},\ }\bibfield  {title} {\bibinfo {title} {Tumbleweeds and
  airborne gravitational noise sources for {LIGO}},\ }\href
  {https://doi.org/10.1088/0264-9381/25/12/125011} {\bibfield  {journal}
  {\bibinfo  {journal} {Classical and Quantum Gravity}\ }\textbf {\bibinfo
  {volume} {25}},\ \bibinfo {pages} {125011} (\bibinfo {year}
  {2008})}\BibitemShut {NoStop}%
\bibitem [{\citenamefont {Harms}(2019)}]{Harms_2019}%
  \BibitemOpen
  \bibfield  {author} {\bibinfo {author} {\bibfnamefont {J.}~\bibnamefont
  {Harms}},\ }\bibfield  {title} {\bibinfo {title} {Terrestrial gravity
  fluctuations},\ }\bibfield  {journal} {\bibinfo  {journal} {Living Reviews in
  Relativity}\ }\textbf {\bibinfo {volume} {22}},\ \href
  {https://doi.org/10.1007/s41114-019-0022-2} {10.1007/s41114-019-0022-2}
  (\bibinfo {year} {2019})\BibitemShut {NoStop}%
\bibitem [{\citenamefont {Baker}\ and\ \citenamefont
  {Thorpe}(2012)}]{Baker:2012ck}%
  \BibitemOpen
  \bibfield  {author} {\bibinfo {author} {\bibfnamefont {J.~G.}\ \bibnamefont
  {Baker}}\ and\ \bibinfo {author} {\bibfnamefont {J.~I.}\ \bibnamefont
  {Thorpe}},\ }\bibfield  {title} {\bibinfo {title} {{Comparison of Atom
  Interferometers and Light Interferometers as Space-Based Gravitational Wave
  Detectors}},\ }\href {https://doi.org/10.1103/PhysRevLett.108.211101}
  {\bibfield  {journal} {\bibinfo  {journal} {Phys. Rev. Lett.}\ }\textbf
  {\bibinfo {volume} {108}},\ \bibinfo {pages} {211101} (\bibinfo {year}
  {2012})},\ \Eprint {https://arxiv.org/abs/1201.5656} {arXiv:1201.5656
  [gr-qc]} \BibitemShut {NoStop}%
\bibitem [{\citenamefont {Vetrano}\ and\ \citenamefont
  {Vicer\'e}(2013)}]{Vetrano:2013qqa}%
  \BibitemOpen
  \bibfield  {author} {\bibinfo {author} {\bibfnamefont {F.}~\bibnamefont
  {Vetrano}}\ and\ \bibinfo {author} {\bibfnamefont {A.}~\bibnamefont
  {Vicer\'e}},\ }\bibfield  {title} {\bibinfo {title} {{Newtonian noise limit
  in atom interferometers for gravitational wave detection}},\ }\href
  {https://doi.org/10.1140/epjc/s10052-013-2590-8} {\bibfield  {journal}
  {\bibinfo  {journal} {Eur. Phys. J. C}\ }\textbf {\bibinfo {volume} {73}},\
  \bibinfo {pages} {2590} (\bibinfo {year} {2013})},\ \Eprint
  {https://arxiv.org/abs/1304.1702} {arXiv:1304.1702 [gr-qc]} \BibitemShut
  {NoStop}%
\bibitem [{\citenamefont {Junca}\ \emph {et~al.}(2019)\citenamefont {Junca}
  \emph {et~al.}}]{MIGAconsortium:2019efk}%
  \BibitemOpen
  \bibfield  {author} {\bibinfo {author} {\bibfnamefont {J.}~\bibnamefont
  {Junca}} \emph {et~al.} (\bibinfo {collaboration} {MIGA consortium}),\
  }\bibfield  {title} {\bibinfo {title} {{Characterizing Earth gravity field
  fluctuations with the MIGA antenna for future Gravitational Wave
  detectors}},\ }\href {https://doi.org/10.1103/PhysRevD.99.104026} {\bibfield
  {journal} {\bibinfo  {journal} {Phys. Rev. D}\ }\textbf {\bibinfo {volume}
  {99}},\ \bibinfo {pages} {104026} (\bibinfo {year} {2019})},\ \Eprint
  {https://arxiv.org/abs/1902.05337} {arXiv:1902.05337 [physics.atom-ph]}
  \BibitemShut {NoStop}%
\bibitem [{\citenamefont {Mitchell}\ \emph {et~al.}(2022)\citenamefont
  {Mitchell}, \citenamefont {Kovachy}, \citenamefont {Hahn}, \citenamefont
  {Adamson},\ and\ \citenamefont {Chattopadhyay}}]{Mitchell:2022zbp}%
  \BibitemOpen
  \bibfield  {author} {\bibinfo {author} {\bibfnamefont {J.~T.}\ \bibnamefont
  {Mitchell}}, \bibinfo {author} {\bibfnamefont {T.}~\bibnamefont {Kovachy}},
  \bibinfo {author} {\bibfnamefont {S.}~\bibnamefont {Hahn}}, \bibinfo {author}
  {\bibfnamefont {P.}~\bibnamefont {Adamson}},\ and\ \bibinfo {author}
  {\bibfnamefont {S.}~\bibnamefont {Chattopadhyay}},\ }\bibfield  {title}
  {\bibinfo {title} {{MAGIS-100 environmental characterization and noise
  analysis}},\ }\href {https://doi.org/10.1088/1748-0221/17/01/P01007}
  {\bibfield  {journal} {\bibinfo  {journal} {JINST}\ }\textbf {\bibinfo
  {volume} {17}}\bibfield  {number} {\bibinfo  {number} { (01)},\ \bibinfo
  {pages} {P01007}},\ }\bibinfo {note} {[Erratum: JINST 17, E02001 (2022)]},\
  \Eprint {https://arxiv.org/abs/2202.04763} {arXiv:2202.04763
  [physics.atom-ph]} \BibitemShut {NoStop}%
\bibitem [{\citenamefont {Yu}\ and\ \citenamefont
  {Tinto}(2011{\natexlab{b}})}]{Yu2011}%
  \BibitemOpen
  \bibfield  {author} {\bibinfo {author} {\bibfnamefont {N.}~\bibnamefont
  {Yu}}\ and\ \bibinfo {author} {\bibfnamefont {M.}~\bibnamefont {Tinto}},\
  }\bibfield  {title} {\bibinfo {title} {Gravitational wave detection with
  single-laser atom interferometers},\ }\href
  {https://doi.org/10.1007/s10714-010-1055-8} {\bibfield  {journal} {\bibinfo
  {journal} {General Relativity and Gravitation}\ }\textbf {\bibinfo {volume}
  {43}},\ \bibinfo {pages} {1943} (\bibinfo {year}
  {2011}{\natexlab{b}})}\BibitemShut {NoStop}%
\bibitem [{\citenamefont {Graham}\ \emph {et~al.}(2013)\citenamefont {Graham},
  \citenamefont {Hogan}, \citenamefont {Kasevich},\ and\ \citenamefont
  {Rajendran}}]{Graham:2012sy}%
  \BibitemOpen
  \bibfield  {author} {\bibinfo {author} {\bibfnamefont {P.~W.}\ \bibnamefont
  {Graham}}, \bibinfo {author} {\bibfnamefont {J.~M.}\ \bibnamefont {Hogan}},
  \bibinfo {author} {\bibfnamefont {M.~A.}\ \bibnamefont {Kasevich}},\ and\
  \bibinfo {author} {\bibfnamefont {S.}~\bibnamefont {Rajendran}},\ }\bibfield
  {title} {\bibinfo {title} {{A New Method for Gravitational Wave Detection
  with Atomic Sensors}},\ }\href
  {https://doi.org/10.1103/PhysRevLett.110.171102} {\bibfield  {journal}
  {\bibinfo  {journal} {Phys. Rev. Lett.}\ }\textbf {\bibinfo {volume} {110}},\
  \bibinfo {pages} {171102} (\bibinfo {year} {2013})},\ \Eprint
  {https://arxiv.org/abs/1206.0818} {arXiv:1206.0818 [quant-ph]} \BibitemShut
  {NoStop}%
\bibitem [{\citenamefont {Hu}\ \emph {et~al.}(2017)\citenamefont {Hu},
  \citenamefont {Poli}, \citenamefont {Salvi},\ and\ \citenamefont
  {Tino}}]{Hu:2017zft}%
  \BibitemOpen
  \bibfield  {author} {\bibinfo {author} {\bibfnamefont {L.}~\bibnamefont
  {Hu}}, \bibinfo {author} {\bibfnamefont {N.}~\bibnamefont {Poli}}, \bibinfo
  {author} {\bibfnamefont {L.}~\bibnamefont {Salvi}},\ and\ \bibinfo {author}
  {\bibfnamefont {G.~M.}\ \bibnamefont {Tino}},\ }\bibfield  {title} {\bibinfo
  {title} {{Atom interferometry with the Sr optical clock transition}},\ }\href
  {https://doi.org/10.1103/PhysRevLett.119.263601} {\bibfield  {journal}
  {\bibinfo  {journal} {Phys. Rev. Lett.}\ }\textbf {\bibinfo {volume} {119}},\
  \bibinfo {pages} {263601} (\bibinfo {year} {2017})},\ \Eprint
  {https://arxiv.org/abs/1708.05116} {arXiv:1708.05116 [physics.atom-ph]}
  \BibitemShut {NoStop}%
\bibitem [{\citenamefont {Hu}\ \emph {et~al.}(2020)\citenamefont {Hu},
  \citenamefont {Wang}, \citenamefont {Salvi}, \citenamefont {Tinsley},
  \citenamefont {Tino},\ and\ \citenamefont {Poli}}]{Hu:2019uey}%
  \BibitemOpen
  \bibfield  {author} {\bibinfo {author} {\bibfnamefont {L.}~\bibnamefont
  {Hu}}, \bibinfo {author} {\bibfnamefont {E.}~\bibnamefont {Wang}}, \bibinfo
  {author} {\bibfnamefont {L.}~\bibnamefont {Salvi}}, \bibinfo {author}
  {\bibfnamefont {J.~N.}\ \bibnamefont {Tinsley}}, \bibinfo {author}
  {\bibfnamefont {G.~M.}\ \bibnamefont {Tino}},\ and\ \bibinfo {author}
  {\bibfnamefont {N.}~\bibnamefont {Poli}},\ }\bibfield  {title} {\bibinfo
  {title} {{Sr atom interferometry with the optical clock transition as a
  gravimeter and a gravity gradiometer}},\ }\href
  {https://doi.org/10.1088/1361-6382/ab4d18} {\bibfield  {journal} {\bibinfo
  {journal} {Class. Quant. Grav.}\ }\textbf {\bibinfo {volume} {37}},\ \bibinfo
  {pages} {014001} (\bibinfo {year} {2020})},\ \Eprint
  {https://arxiv.org/abs/1907.10537} {arXiv:1907.10537 [physics.atom-ph]}
  \BibitemShut {NoStop}%
\bibitem [{\citenamefont {McGuirk}\ \emph {et~al.}(2000)\citenamefont
  {McGuirk}, \citenamefont {Snadden},\ and\ \citenamefont
  {Kasevich}}]{McGuirk:2000zz}%
  \BibitemOpen
  \bibfield  {author} {\bibinfo {author} {\bibfnamefont {J.~M.}\ \bibnamefont
  {McGuirk}}, \bibinfo {author} {\bibfnamefont {M.~J.}\ \bibnamefont
  {Snadden}},\ and\ \bibinfo {author} {\bibfnamefont {M.~A.}\ \bibnamefont
  {Kasevich}},\ }\bibfield  {title} {\bibinfo {title} {{Large Area Light-Pulse
  Atom Interferometry}},\ }\href {https://doi.org/10.1103/PhysRevLett.85.4498}
  {\bibfield  {journal} {\bibinfo  {journal} {Phys. Rev. Lett.}\ }\textbf
  {\bibinfo {volume} {85}},\ \bibinfo {pages} {4498} (\bibinfo {year}
  {2000})}\BibitemShut {NoStop}%
\bibitem [{\citenamefont {Chiow}\ \emph {et~al.}(2011)\citenamefont {Chiow},
  \citenamefont {Kovachy}, \citenamefont {Chien},\ and\ \citenamefont
  {Kasevich}}]{Chiow:2011zz}%
  \BibitemOpen
  \bibfield  {author} {\bibinfo {author} {\bibfnamefont {S.-w.}\ \bibnamefont
  {Chiow}}, \bibinfo {author} {\bibfnamefont {T.}~\bibnamefont {Kovachy}},
  \bibinfo {author} {\bibfnamefont {H.-C.}\ \bibnamefont {Chien}},\ and\
  \bibinfo {author} {\bibfnamefont {M.~A.}\ \bibnamefont {Kasevich}},\
  }\bibfield  {title} {\bibinfo {title} {{102 h-bar k Large Area Atom
  Interferometers}},\ }\href {https://doi.org/10.1103/PhysRevLett.107.130403}
  {\bibfield  {journal} {\bibinfo  {journal} {Phys. Rev. Lett.}\ }\textbf
  {\bibinfo {volume} {107}},\ \bibinfo {pages} {130403} (\bibinfo {year}
  {2011})}\BibitemShut {NoStop}%
\bibitem [{\citenamefont {Rudolph}\ \emph {et~al.}(2020)\citenamefont
  {Rudolph}, \citenamefont {Wilkason} \emph {et~al.}}]{Rudolph_2020}%
  \BibitemOpen
  \bibfield  {author} {\bibinfo {author} {\bibfnamefont {J.}~\bibnamefont
  {Rudolph}}, \bibinfo {author} {\bibfnamefont {T.}~\bibnamefont {Wilkason}},
  \emph {et~al.},\ }\bibfield  {title} {\bibinfo {title} {Large momentum
  transfer clock atom interferometry on the 689 nm intercombination line of
  strontium},\ }\href {https://doi.org/10.1103/PhysRevLett.124.083604}
  {\bibfield  {journal} {\bibinfo  {journal} {Phys. Rev. Lett.}\ }\textbf
  {\bibinfo {volume} {124}},\ \bibinfo {pages} {083604} (\bibinfo {year}
  {2020})}\BibitemShut {NoStop}%
\bibitem [{\citenamefont {Wilkason}\ \emph {et~al.}(2022)\citenamefont
  {Wilkason}, \citenamefont {Nantel} \emph {et~al.}}]{Wilkason:2022yej}%
  \BibitemOpen
  \bibfield  {author} {\bibinfo {author} {\bibfnamefont {T.}~\bibnamefont
  {Wilkason}}, \bibinfo {author} {\bibfnamefont {M.}~\bibnamefont {Nantel}},
  \emph {et~al.},\ }\bibfield  {title} {\bibinfo {title} {{Atom Interferometry
  with Floquet Atom Optics}},\ }\href
  {https://doi.org/10.1103/PhysRevLett.129.183202} {\bibfield  {journal}
  {\bibinfo  {journal} {Phys. Rev. Lett.}\ }\textbf {\bibinfo {volume} {129}},\
  \bibinfo {pages} {183202} (\bibinfo {year} {2022})},\ \Eprint
  {https://arxiv.org/abs/2205.06965} {arXiv:2205.06965 [physics.atom-ph]}
  \BibitemShut {NoStop}%
\bibitem [{\citenamefont {Hui}(2021)}]{Hui:2021tkt}%
  \BibitemOpen
  \bibfield  {author} {\bibinfo {author} {\bibfnamefont {L.}~\bibnamefont
  {Hui}},\ }\bibfield  {title} {\bibinfo {title} {{Wave Dark Matter}},\ }\href
  {https://doi.org/10.1146/annurev-astro-120920-010024} {\bibfield  {journal}
  {\bibinfo  {journal} {Ann. Rev. Astron. Astrophys.}\ }\textbf {\bibinfo
  {volume} {59}},\ \bibinfo {pages} {247} (\bibinfo {year} {2021})},\ \Eprint
  {https://arxiv.org/abs/2101.11735} {arXiv:2101.11735 [astro-ph.CO]}
  \BibitemShut {NoStop}%
\bibitem [{\citenamefont {Armaleo}\ \emph {et~al.}(2021)\citenamefont
  {Armaleo}, \citenamefont {Nacir},\ and\ \citenamefont
  {Urban}}]{Armaleo_2021}%
  \BibitemOpen
  \bibfield  {author} {\bibinfo {author} {\bibfnamefont {J.~M.}\ \bibnamefont
  {Armaleo}}, \bibinfo {author} {\bibfnamefont {D.~L.}\ \bibnamefont {Nacir}},\
  and\ \bibinfo {author} {\bibfnamefont {F.~R.}\ \bibnamefont {Urban}},\
  }\bibfield  {title} {\bibinfo {title} {Searching for spin-2 {ULDM} with
  gravitational waves interferometers},\ }\href
  {https://doi.org/10.1088/1475-7516/2021/04/053} {\bibfield  {journal}
  {\bibinfo  {journal} {Journal of Cosmology and Astroparticle Physics}\
  }\textbf {\bibinfo {volume} {2021}}\bibinfo  {number} { (04)},\ \bibinfo
  {pages} {053}}\BibitemShut {NoStop}%
\bibitem [{\citenamefont {Graham}\ \emph {et~al.}(2018)\citenamefont {Graham},
  \citenamefont {Kaplan} \emph {et~al.}}]{Graham:2017ivz}%
  \BibitemOpen
\bibfield  {number} {  }\bibfield  {author} {\bibinfo {author} {\bibfnamefont
  {P.~W.}\ \bibnamefont {Graham}}, \bibinfo {author} {\bibfnamefont {D.~E.}\
  \bibnamefont {Kaplan}}, \emph {et~al.},\ }\bibfield  {title} {\bibinfo
  {title} {{Spin Precession Experiments for Light Axionic Dark Matter}},\
  }\href {https://doi.org/10.1103/PhysRevD.97.055006} {\bibfield  {journal}
  {\bibinfo  {journal} {Phys. Rev. D}\ }\textbf {\bibinfo {volume} {97}},\
  \bibinfo {pages} {055006} (\bibinfo {year} {2018})},\ \Eprint
  {https://arxiv.org/abs/1709.07852} {arXiv:1709.07852 [hep-ph]} \BibitemShut
  {NoStop}%
\bibitem [{\citenamefont {Read}(2014)}]{Read:2014qva}%
  \BibitemOpen
  \bibfield  {author} {\bibinfo {author} {\bibfnamefont {J.~I.}\ \bibnamefont
  {Read}},\ }\bibfield  {title} {\bibinfo {title} {{The Local Dark Matter
  Density}},\ }\href {https://doi.org/10.1088/0954-3899/41/6/063101} {\bibfield
   {journal} {\bibinfo  {journal} {J. Phys. G}\ }\textbf {\bibinfo {volume}
  {41}},\ \bibinfo {pages} {063101} (\bibinfo {year} {2014})},\ \Eprint
  {https://arxiv.org/abs/1404.1938} {arXiv:1404.1938 [astro-ph.GA]}
  \BibitemShut {NoStop}%
\bibitem [{\citenamefont {Kaplan}\ and\ \citenamefont
  {Wise}(2000)}]{Kaplan_2000}%
  \BibitemOpen
  \bibfield  {author} {\bibinfo {author} {\bibfnamefont {D.~B.}\ \bibnamefont
  {Kaplan}}\ and\ \bibinfo {author} {\bibfnamefont {M.~B.}\ \bibnamefont
  {Wise}},\ }\bibfield  {title} {\bibinfo {title} {{Couplings of a light
  dilaton and violations of the equivalence principle}},\ }\href
  {https://doi.org/10.1088/1126-6708/2000/08/037} {\bibfield  {journal}
  {\bibinfo  {journal} {JHEP}\ }\textbf {\bibinfo {volume} {2000}}\bibfield
  {number} {\bibinfo  {number} { (08)},\ \bibinfo {pages} {037}},\ }\Eprint
  {https://arxiv.org/abs/hep-ph/0008116} {arXiv:hep-ph/0008116} \BibitemShut
  {NoStop}%
\bibitem [{\citenamefont {Damour}\ and\ \citenamefont
  {Donoghue}(2010)}]{Damour:2010rp}%
  \BibitemOpen
  \bibfield  {author} {\bibinfo {author} {\bibfnamefont {T.}~\bibnamefont
  {Damour}}\ and\ \bibinfo {author} {\bibfnamefont {J.~F.}\ \bibnamefont
  {Donoghue}},\ }\bibfield  {title} {\bibinfo {title} {{Equivalence Principle
  Violations and Couplings of a Light Dilaton}},\ }\href
  {https://doi.org/10.1103/PhysRevD.82.084033} {\bibfield  {journal} {\bibinfo
  {journal} {Phys. Rev. D}\ }\textbf {\bibinfo {volume} {82}},\ \bibinfo
  {pages} {084033} (\bibinfo {year} {2010})},\ \Eprint
  {https://arxiv.org/abs/1007.2792} {arXiv:1007.2792 [gr-qc]} \BibitemShut
  {NoStop}%
\bibitem [{\citenamefont {Stadnik}\ and\ \citenamefont
  {Flambaum}(2015{\natexlab{a}})}]{Stadnik:2015kia}%
  \BibitemOpen
  \bibfield  {author} {\bibinfo {author} {\bibfnamefont {Y.~V.}\ \bibnamefont
  {Stadnik}}\ and\ \bibinfo {author} {\bibfnamefont {V.~V.}\ \bibnamefont
  {Flambaum}},\ }\bibfield  {title} {\bibinfo {title} {{Can dark matter induce
  cosmological evolution of the fundamental constants of Nature?}},\ }\href
  {https://doi.org/10.1103/PhysRevLett.115.201301} {\bibfield  {journal}
  {\bibinfo  {journal} {Phys. Rev. Lett.}\ }\textbf {\bibinfo {volume} {115}},\
  \bibinfo {pages} {201301} (\bibinfo {year} {2015}{\natexlab{a}})},\ \Eprint
  {https://arxiv.org/abs/1503.08540} {arXiv:1503.08540 [astro-ph.CO]}
  \BibitemShut {NoStop}%
\bibitem [{\citenamefont {Hees}\ \emph {et~al.}(2018)\citenamefont {Hees},
  \citenamefont {Minazzoli} \emph {et~al.}}]{Hees:2018fpg}%
  \BibitemOpen
  \bibfield  {author} {\bibinfo {author} {\bibfnamefont {A.}~\bibnamefont
  {Hees}}, \bibinfo {author} {\bibfnamefont {O.}~\bibnamefont {Minazzoli}},
  \emph {et~al.},\ }\bibfield  {title} {\bibinfo {title} {{Violation of the
  equivalence principle from light scalar dark matter}},\ }\href
  {https://doi.org/10.1103/PhysRevD.98.064051} {\bibfield  {journal} {\bibinfo
  {journal} {Phys. Rev. D}\ }\textbf {\bibinfo {volume} {98}},\ \bibinfo
  {pages} {064051} (\bibinfo {year} {2018})},\ \Eprint
  {https://arxiv.org/abs/1807.04512} {arXiv:1807.04512 [gr-qc]} \BibitemShut
  {NoStop}%
\bibitem [{\citenamefont {Bouley}\ \emph {et~al.}(2023)\citenamefont {Bouley},
  \citenamefont {S\o{}rensen},\ and\ \citenamefont {Yu}}]{Bouley:2022eer}%
  \BibitemOpen
  \bibfield  {author} {\bibinfo {author} {\bibfnamefont {T.}~\bibnamefont
  {Bouley}}, \bibinfo {author} {\bibfnamefont {P.}~\bibnamefont
  {S\o{}rensen}},\ and\ \bibinfo {author} {\bibfnamefont {T.-T.}\ \bibnamefont
  {Yu}},\ }\bibfield  {title} {\bibinfo {title} {{Constraints on ultralight
  scalar dark matter with quadratic couplings}},\ }\href
  {https://doi.org/10.1007/JHEP03(2023)104} {\bibfield  {journal} {\bibinfo
  {journal} {JHEP}\ }\textbf {\bibinfo {volume} {2023}}\bibfield  {number}
  {\bibinfo  {number} { (03)},\ \bibinfo {pages} {104}},\ }\Eprint
  {https://arxiv.org/abs/2211.09826} {arXiv:2211.09826 [hep-ph]} \BibitemShut
  {NoStop}%
\bibitem [{\citenamefont {Stadnik}\ and\ \citenamefont
  {Flambaum}(2015{\natexlab{b}})}]{Stadnik_2014}%
  \BibitemOpen
  \bibfield  {author} {\bibinfo {author} {\bibfnamefont {Y.~V.}\ \bibnamefont
  {Stadnik}}\ and\ \bibinfo {author} {\bibfnamefont {V.~V.}\ \bibnamefont
  {Flambaum}},\ }\bibfield  {title} {\bibinfo {title} {{Searching for dark
  matter and variation of fundamental constants with laser and maser
  interferometry}},\ }\href {https://doi.org/10.1103/PhysRevLett.114.161301}
  {\bibfield  {journal} {\bibinfo  {journal} {Phys. Rev. Lett.}\ }\textbf
  {\bibinfo {volume} {114}},\ \bibinfo {pages} {161301} (\bibinfo {year}
  {2015}{\natexlab{b}})},\ \Eprint {https://arxiv.org/abs/1412.7801}
  {arXiv:1412.7801 [hep-ph]} \BibitemShut {NoStop}%
\bibitem [{\citenamefont {Derevianko}(2018)}]{Derevianko:2016vpm}%
  \BibitemOpen
  \bibfield  {author} {\bibinfo {author} {\bibfnamefont {A.}~\bibnamefont
  {Derevianko}},\ }\bibfield  {title} {\bibinfo {title} {{Detecting dark-matter
  waves with a network of precision-measurement tools}},\ }\href
  {https://doi.org/10.1103/PhysRevA.97.042506} {\bibfield  {journal} {\bibinfo
  {journal} {Phys. Rev. A}\ }\textbf {\bibinfo {volume} {97}},\ \bibinfo
  {pages} {042506} (\bibinfo {year} {2018})},\ \Eprint
  {https://arxiv.org/abs/1605.09717} {arXiv:1605.09717 [physics.atom-ph]}
  \BibitemShut {NoStop}%
\bibitem [{\citenamefont {Foster}\ \emph {et~al.}(2018)\citenamefont {Foster},
  \citenamefont {Rodd},\ and\ \citenamefont {Safdi}}]{Foster:2017hbq}%
  \BibitemOpen
  \bibfield  {author} {\bibinfo {author} {\bibfnamefont {J.~W.}\ \bibnamefont
  {Foster}}, \bibinfo {author} {\bibfnamefont {N.~L.}\ \bibnamefont {Rodd}},\
  and\ \bibinfo {author} {\bibfnamefont {B.~R.}\ \bibnamefont {Safdi}},\
  }\bibfield  {title} {\bibinfo {title} {{Revealing the Dark Matter Halo with
  Axion Direct Detection}},\ }\href
  {https://doi.org/10.1103/PhysRevD.97.123006} {\bibfield  {journal} {\bibinfo
  {journal} {Phys. Rev. D}\ }\textbf {\bibinfo {volume} {97}},\ \bibinfo
  {pages} {123006} (\bibinfo {year} {2018})},\ \Eprint
  {https://arxiv.org/abs/1711.10489} {arXiv:1711.10489 [astro-ph.CO]}
  \BibitemShut {NoStop}%
\bibitem [{\citenamefont {Centers}\ \emph {et~al.}(2021)\citenamefont {Centers}
  \emph {et~al.}}]{Centers:2019dyn}%
  \BibitemOpen
  \bibfield  {author} {\bibinfo {author} {\bibfnamefont {G.~P.}\ \bibnamefont
  {Centers}} \emph {et~al.},\ }\bibfield  {title} {\bibinfo {title}
  {{Stochastic fluctuations of bosonic dark matter}},\ }\href
  {https://doi.org/10.1038/s41467-021-27632-7} {\bibfield  {journal} {\bibinfo
  {journal} {Nature Commun.}\ }\textbf {\bibinfo {volume} {12}},\ \bibinfo
  {pages} {7321} (\bibinfo {year} {2021})},\ \Eprint
  {https://arxiv.org/abs/1905.13650} {arXiv:1905.13650 [astro-ph.CO]}
  \BibitemShut {NoStop}%
\bibitem [{\citenamefont {Nakatsuka}\ \emph {et~al.}(2022)\citenamefont
  {Nakatsuka}, \citenamefont {Morisaki} \emph {et~al.}}]{Nakatsuka:2022gaf}%
  \BibitemOpen
  \bibfield  {author} {\bibinfo {author} {\bibfnamefont {H.}~\bibnamefont
  {Nakatsuka}}, \bibinfo {author} {\bibfnamefont {S.}~\bibnamefont {Morisaki}},
  \emph {et~al.},\ }\bibfield  {title} {\bibinfo {title} {{Stochastic effects
  on observation of ultralight bosonic dark matter}}\ }(\bibinfo {year}
  {2022})\ \Eprint {https://arxiv.org/abs/2205.02960} {arXiv:2205.02960
  [astro-ph.CO]} \BibitemShut {NoStop}%
\bibitem [{\citenamefont {Angstmann}\ \emph {et~al.}(2004)\citenamefont
  {Angstmann}, \citenamefont {Dzuba},\ and\ \citenamefont
  {Flambaum}}]{Angstmann:2004zz}%
  \BibitemOpen
  \bibfield  {author} {\bibinfo {author} {\bibfnamefont {E.~J.}\ \bibnamefont
  {Angstmann}}, \bibinfo {author} {\bibfnamefont {V.~A.}\ \bibnamefont
  {Dzuba}},\ and\ \bibinfo {author} {\bibfnamefont {V.~V.}\ \bibnamefont
  {Flambaum}},\ }\bibfield  {title} {\bibinfo {title} {{Relativistic effects in
  two valence electron atoms and ions and the search for variation of the fine
  structure constant}},\ }\href {https://doi.org/10.1103/PhysRevA.70.014102}
  {\bibfield  {journal} {\bibinfo  {journal} {Phys. Rev. A}\ }\textbf {\bibinfo
  {volume} {70}},\ \bibinfo {pages} {014102} (\bibinfo {year} {2004})},\
  \Eprint {https://arxiv.org/abs/physics/0404042} {arXiv:physics/0404042}
  \BibitemShut {NoStop}%
\bibitem [{\citenamefont {Touboul}\ \emph {et~al.}(2022)\citenamefont {Touboul}
  \emph {et~al.}}]{MICROSCOPE:2022doy}%
  \BibitemOpen
  \bibfield  {author} {\bibinfo {author} {\bibfnamefont {P.}~\bibnamefont
  {Touboul}} \emph {et~al.} (\bibinfo {collaboration} {MICROSCOPE}),\
  }\bibfield  {title} {\bibinfo {title} {{MICROSCOPE Mission: Final Results of
  the Test of the Equivalence Principle}},\ }\href
  {https://doi.org/10.1103/PhysRevLett.129.121102} {\bibfield  {journal}
  {\bibinfo  {journal} {Phys. Rev. Lett.}\ }\textbf {\bibinfo {volume} {129}},\
  \bibinfo {pages} {121102} (\bibinfo {year} {2022})},\ \Eprint
  {https://arxiv.org/abs/2209.15487} {arXiv:2209.15487 [gr-qc]} \BibitemShut
  {NoStop}%
\bibitem [{\citenamefont {Schuster}(1898)}]{schuster1898investigation}%
  \BibitemOpen
  \bibfield  {author} {\bibinfo {author} {\bibfnamefont {A.}~\bibnamefont
  {Schuster}},\ }\bibfield  {title} {\bibinfo {title} {On the investigation of
  hidden periodicities with application to a supposed 26 day period of
  meteorological phenomena},\ }\href@noop {} {\bibfield  {journal} {\bibinfo
  {journal} {Terrestrial Magnetism}\ }\textbf {\bibinfo {volume} {3}},\
  \bibinfo {pages} {13} (\bibinfo {year} {1898})}\BibitemShut {NoStop}%
\bibitem [{\citenamefont {Badurina}\ \emph
  {et~al.}(2023{\natexlab{b}})\citenamefont {Badurina}, \citenamefont
  {Beniwal},\ and\ \citenamefont {McCabe}}]{Badurina:2023wpk}%
  \BibitemOpen
  \bibfield  {author} {\bibinfo {author} {\bibfnamefont {L.}~\bibnamefont
  {Badurina}}, \bibinfo {author} {\bibfnamefont {A.}~\bibnamefont {Beniwal}},\
  and\ \bibinfo {author} {\bibfnamefont {C.}~\bibnamefont {McCabe}},\
  }\bibfield  {title} {\bibinfo {title} {{Super-Nyquist ultralight dark matter
  searches with broadband atom gradiometers}}\ }(\bibinfo {year} {2023})\
  \Eprint {https://arxiv.org/abs/2306.16477} {arXiv:2306.16477 [hep-ph]}
  \BibitemShut {NoStop}%
\bibitem [{\citenamefont {Hogan}\ \emph {et~al.}(2009)\citenamefont {Hogan},
  \citenamefont {Johnson},\ and\ \citenamefont {Kasevich}}]{Hogan_2008}%
  \BibitemOpen
  \bibfield  {author} {\bibinfo {author} {\bibfnamefont {J.}~\bibnamefont
  {Hogan}}, \bibinfo {author} {\bibfnamefont {D.}~\bibnamefont {Johnson}},\
  and\ \bibinfo {author} {\bibfnamefont {M.}~\bibnamefont {Kasevich}},\
  }\bibfield  {title} {\bibinfo {title} {Light-pulse atom interferometry},\
  }in\ \href {https://doi.org/10.48550/ARXIV.0806.3261} {\emph {\bibinfo
  {booktitle} {Atom Optics and Space Physics}}}\ (\bibinfo  {publisher} {IOS
  Press},\ \bibinfo {year} {2009})\ pp.\ \bibinfo {pages} {411--447},\ \Eprint
  {https://arxiv.org/abs/0806.3261} {arXiv:0806.3261 [physics.atom-ph]}
  \BibitemShut {NoStop}%
\bibitem [{\citenamefont {{Pippa Storey}}\ and\ \citenamefont {{Claude
  Cohen-Tannoudji}}(1994)}]{Storey}%
  \BibitemOpen
  \bibfield  {author} {\bibinfo {author} {\bibnamefont {{Pippa Storey}}}\ and\
  \bibinfo {author} {\bibnamefont {{Claude Cohen-Tannoudji}}},\ }\bibfield
  {title} {\bibinfo {title} {The {F}eynman path integral approach to atomic
  interferometry. {A}~tutorial},\ }\href {https://doi.org/10.1051/jp2:1994103}
  {\bibfield  {journal} {\bibinfo  {journal} {J. Phys. II France}\ }\textbf
  {\bibinfo {volume} {4}},\ \bibinfo {pages} {1999} (\bibinfo {year}
  {1994})}\BibitemShut {NoStop}%
\bibitem [{\citenamefont {Peters}\ \emph
  {et~al.}(2001{\natexlab{b}})\citenamefont {Peters}, \citenamefont {Chung},\
  and\ \citenamefont {Chu}}]{peters2001high}%
  \BibitemOpen
  \bibfield  {author} {\bibinfo {author} {\bibfnamefont {A.}~\bibnamefont
  {Peters}}, \bibinfo {author} {\bibfnamefont {K.~Y.}\ \bibnamefont {Chung}},\
  and\ \bibinfo {author} {\bibfnamefont {S.}~\bibnamefont {Chu}},\ }\bibfield
  {title} {\bibinfo {title} {High-precision gravity measurements using atom
  interferometry},\ }\href@noop {} {\bibfield  {journal} {\bibinfo  {journal}
  {Metrologia}\ }\textbf {\bibinfo {volume} {38}},\ \bibinfo {pages} {25}
  (\bibinfo {year} {2001}{\natexlab{b}})}\BibitemShut {NoStop}%
\bibitem [{\citenamefont {Bongs}\ \emph {et~al.}(2006)\citenamefont {Bongs},
  \citenamefont {Launay},\ and\ \citenamefont {Kasevich}}]{Bongs_2002}%
  \BibitemOpen
  \bibfield  {author} {\bibinfo {author} {\bibfnamefont {K.}~\bibnamefont
  {Bongs}}, \bibinfo {author} {\bibfnamefont {R.}~\bibnamefont {Launay}},\ and\
  \bibinfo {author} {\bibfnamefont {M.~A.}\ \bibnamefont {Kasevich}},\
  }\bibfield  {title} {\bibinfo {title} {High-order inertial phase shifts for
  time-domain atom interferometers},\ }\href@noop {} {\bibfield  {journal}
  {\bibinfo  {journal} {Applied Physics B}\ }\textbf {\bibinfo {volume} {84}},\
  \bibinfo {pages} {599} (\bibinfo {year} {2006})},\ \Eprint
  {https://arxiv.org/abs/quant-ph/0204102} {arXiv:quant-ph/0204102 [quant-ph]}
  \BibitemShut {NoStop}%
\bibitem [{\citenamefont {Overstreet}\ \emph
  {et~al.}(2021{\natexlab{a}})\citenamefont {Overstreet}, \citenamefont
  {Asenbaum},\ and\ \citenamefont {Kasevich}}]{Overstreet_2021}%
  \BibitemOpen
  \bibfield  {author} {\bibinfo {author} {\bibfnamefont {C.}~\bibnamefont
  {Overstreet}}, \bibinfo {author} {\bibfnamefont {P.}~\bibnamefont
  {Asenbaum}},\ and\ \bibinfo {author} {\bibfnamefont {M.~A.}\ \bibnamefont
  {Kasevich}},\ }\bibfield  {title} {\bibinfo {title} {{Physically significant
  phase shifts in matter-wave interferometry}},\ }\href
  {https://doi.org/10.1119/10.0002638} {\bibfield  {journal} {\bibinfo
  {journal} {Am. J. Phys.}\ }\textbf {\bibinfo {volume} {89}},\ \bibinfo
  {pages} {324} (\bibinfo {year} {2021}{\natexlab{a}})},\ \Eprint
  {https://arxiv.org/abs/2008.05609} {arXiv:2008.05609 [physics.atom-ph]}
  \BibitemShut {NoStop}%
\bibitem [{\citenamefont {Bertoldi}\ \emph {et~al.}(2019)\citenamefont
  {Bertoldi}, \citenamefont {Minardi},\ and\ \citenamefont
  {Prevedelli}}]{Bertoldi:2018zga}%
  \BibitemOpen
  \bibfield  {author} {\bibinfo {author} {\bibfnamefont {A.}~\bibnamefont
  {Bertoldi}}, \bibinfo {author} {\bibfnamefont {F.}~\bibnamefont {Minardi}},\
  and\ \bibinfo {author} {\bibfnamefont {M.}~\bibnamefont {Prevedelli}},\
  }\bibfield  {title} {\bibinfo {title} {{Phase shift in atom interferometers:
  Corrections for nonquadratic potentials and finite-duration laser pulses}},\
  }\href {https://doi.org/10.1103/PhysRevA.99.033619} {\bibfield  {journal}
  {\bibinfo  {journal} {Phys. Rev. A}\ }\textbf {\bibinfo {volume} {99}},\
  \bibinfo {pages} {033619} (\bibinfo {year} {2019})},\ \Eprint
  {https://arxiv.org/abs/1812.11890} {arXiv:1812.11890 [quant-ph]} \BibitemShut
  {NoStop}%
\bibitem [{\citenamefont {Ufrecht}\ and\ \citenamefont
  {Giese}(2020)}]{Ufrecht:2020vcr}%
  \BibitemOpen
  \bibfield  {author} {\bibinfo {author} {\bibfnamefont {C.}~\bibnamefont
  {Ufrecht}}\ and\ \bibinfo {author} {\bibfnamefont {E.}~\bibnamefont
  {Giese}},\ }\bibfield  {title} {\bibinfo {title} {{Perturbative operator
  approach to high-precision light-pulse atom interferometry}},\ }\href
  {https://doi.org/10.1103/PhysRevA.101.053615} {\bibfield  {journal} {\bibinfo
   {journal} {Phys. Rev. A}\ }\textbf {\bibinfo {volume} {101}},\ \bibinfo
  {pages} {053615} (\bibinfo {year} {2020})},\ \Eprint
  {https://arxiv.org/abs/2003.02042} {arXiv:2003.02042 [quant-ph]} \BibitemShut
  {NoStop}%
\bibitem [{\citenamefont {Overstreet}\ \emph
  {et~al.}(2021{\natexlab{b}})\citenamefont {Overstreet}, \citenamefont
  {Asenbaum}, \citenamefont {Curti}, \citenamefont {Kim},\ and\ \citenamefont
  {Kasevich}}]{Overstreet:2021hea}%
  \BibitemOpen
  \bibfield  {author} {\bibinfo {author} {\bibfnamefont {C.}~\bibnamefont
  {Overstreet}}, \bibinfo {author} {\bibfnamefont {P.}~\bibnamefont
  {Asenbaum}}, \bibinfo {author} {\bibfnamefont {J.}~\bibnamefont {Curti}},
  \bibinfo {author} {\bibfnamefont {M.}~\bibnamefont {Kim}},\ and\ \bibinfo
  {author} {\bibfnamefont {M.~A.}\ \bibnamefont {Kasevich}},\ }\bibfield
  {title} {\bibinfo {title} {{Observation of a gravitational Aharonov-Bohm
  effect}},\ }\href {https://doi.org/10.1126/science.abl7152} {\bibfield
  {journal} {\bibinfo  {journal} {Science}\ }\textbf {\bibinfo {volume}
  {375}},\ \bibinfo {pages} {abl7152} (\bibinfo {year}
  {2021}{\natexlab{b}})}\BibitemShut {NoStop}%
\bibitem [{\citenamefont {Roura}(2017)}]{Roura:2015xsa}%
  \BibitemOpen
  \bibfield  {author} {\bibinfo {author} {\bibfnamefont {A.}~\bibnamefont
  {Roura}},\ }\bibfield  {title} {\bibinfo {title} {{Circumventing Heisenberg's
  uncertainty principle in atom interferometry tests of the equivalence
  principle}},\ }\href {https://doi.org/10.1103/PhysRevLett.118.160401}
  {\bibfield  {journal} {\bibinfo  {journal} {Phys. Rev. Lett.}\ }\textbf
  {\bibinfo {volume} {118}},\ \bibinfo {pages} {160401} (\bibinfo {year}
  {2017})},\ \Eprint {https://arxiv.org/abs/1509.08098} {arXiv:1509.08098
  [physics.atom-ph]} \BibitemShut {NoStop}%
\bibitem [{\citenamefont {D'Amico}\ \emph {et~al.}(2017)\citenamefont
  {D'Amico}, \citenamefont {Rosi}, \citenamefont {Zhan}, \citenamefont
  {Cacciapuoti}, \citenamefont {Fattori},\ and\ \citenamefont
  {Tino}}]{PhysRevLett.119.253201}%
  \BibitemOpen
  \bibfield  {author} {\bibinfo {author} {\bibfnamefont {G.}~\bibnamefont
  {D'Amico}}, \bibinfo {author} {\bibfnamefont {G.}~\bibnamefont {Rosi}},
  \bibinfo {author} {\bibfnamefont {S.}~\bibnamefont {Zhan}}, \bibinfo {author}
  {\bibfnamefont {L.}~\bibnamefont {Cacciapuoti}}, \bibinfo {author}
  {\bibfnamefont {M.}~\bibnamefont {Fattori}},\ and\ \bibinfo {author}
  {\bibfnamefont {G.~M.}\ \bibnamefont {Tino}},\ }\bibfield  {title} {\bibinfo
  {title} {Canceling the gravity gradient phase shift in atom interferometry},\
  }\href {https://doi.org/10.1103/PhysRevLett.119.253201} {\bibfield  {journal}
  {\bibinfo  {journal} {Phys. Rev. Lett.}\ }\textbf {\bibinfo {volume} {119}},\
  \bibinfo {pages} {253201} (\bibinfo {year} {2017})}\BibitemShut {NoStop}%
\bibitem [{\citenamefont {{Deeming}}(1975)}]{Deeming1975}%
  \BibitemOpen
  \bibfield  {author} {\bibinfo {author} {\bibfnamefont {T.~J.}\ \bibnamefont
  {{Deeming}}},\ }\bibfield  {title} {\bibinfo {title} {{Fourier Analysis with
  Unequally-Spaced Data}},\ }\href {https://doi.org/10.1007/BF00681947}
  {\bibfield  {journal} {\bibinfo  {journal} {Astrophysics and Space Science}\
  }\textbf {\bibinfo {volume} {36}},\ \bibinfo {pages} {137} (\bibinfo {year}
  {1975})}\BibitemShut {NoStop}%
\bibitem [{\citenamefont {{Carbonell}}\ \emph {et~al.}(1992)\citenamefont
  {{Carbonell}}, \citenamefont {{Oliver}},\ and\ \citenamefont
  {{Ballester}}}]{Carbonell1992}%
  \BibitemOpen
  \bibfield  {author} {\bibinfo {author} {\bibfnamefont {M.}~\bibnamefont
  {{Carbonell}}}, \bibinfo {author} {\bibfnamefont {R.}~\bibnamefont
  {{Oliver}}},\ and\ \bibinfo {author} {\bibfnamefont {J.~L.}\ \bibnamefont
  {{Ballester}}},\ }\bibfield  {title} {\bibinfo {title} {{Power spectra of
  gapped time series - A comparison of several methods}},\ }\href@noop {}
  {\bibfield  {journal} {\bibinfo  {journal} {Astronomy and Astrophysics}\
  }\textbf {\bibinfo {volume} {264}},\ \bibinfo {pages} {350} (\bibinfo {year}
  {1992})}\BibitemShut {NoStop}%
\bibitem [{\citenamefont {Huppert}\ \emph {et~al.}(2009)\citenamefont
  {Huppert}, \citenamefont {Diamond}, \citenamefont {Franceschini},\ and\
  \citenamefont {Boas}}]{huppert2009homer}%
  \BibitemOpen
  \bibfield  {author} {\bibinfo {author} {\bibfnamefont {T.~J.}\ \bibnamefont
  {Huppert}}, \bibinfo {author} {\bibfnamefont {S.~G.}\ \bibnamefont
  {Diamond}}, \bibinfo {author} {\bibfnamefont {M.~A.}\ \bibnamefont
  {Franceschini}},\ and\ \bibinfo {author} {\bibfnamefont {D.~A.}\ \bibnamefont
  {Boas}},\ }\bibfield  {title} {\bibinfo {title} {Homer: a review of
  time-series analysis methods for near-infrared spectroscopy of the brain},\
  }\href@noop {} {\bibfield  {journal} {\bibinfo  {journal} {Applied optics}\
  }\textbf {\bibinfo {volume} {48}},\ \bibinfo {pages} {D280} (\bibinfo {year}
  {2009})}\BibitemShut {NoStop}%
\bibitem [{\citenamefont {Feuerstein}\ \emph {et~al.}(2009)\citenamefont
  {Feuerstein}, \citenamefont {Parker},\ and\ \citenamefont
  {Boutelle}}]{Feuerstein2009}%
  \BibitemOpen
  \bibfield  {author} {\bibinfo {author} {\bibfnamefont {D.}~\bibnamefont
  {Feuerstein}}, \bibinfo {author} {\bibfnamefont {K.~H.}\ \bibnamefont
  {Parker}},\ and\ \bibinfo {author} {\bibfnamefont {M.~G.}\ \bibnamefont
  {Boutelle}},\ }\bibfield  {title} {\bibinfo {title} {Practical methods for
  noise removal: Applications to spikes, nonstationary quasi-periodic noise,
  and baseline drift},\ }\href {https://doi.org/10.1021/ac900161x} {\bibfield
  {journal} {\bibinfo  {journal} {Analytical Chemistry}\ }\textbf {\bibinfo
  {volume} {81}},\ \bibinfo {pages} {4987} (\bibinfo {year}
  {2009})}\BibitemShut {NoStop}%
\bibitem [{\citenamefont {Patton}(2012)}]{PATTON2012}%
  \BibitemOpen
  \bibfield  {author} {\bibinfo {author} {\bibfnamefont {A.~J.}\ \bibnamefont
  {Patton}},\ }\bibfield  {title} {\bibinfo {title} {A review of copula models
  for economic time series},\ }\href
  {https://doi.org/https://doi.org/10.1016/j.jmva.2012.02.021} {\bibfield
  {journal} {\bibinfo  {journal} {Journal of Multivariate Analysis}\ }\textbf
  {\bibinfo {volume} {110}},\ \bibinfo {pages} {4} (\bibinfo {year}
  {2012})}\BibitemShut {NoStop}%
\bibitem [{\citenamefont {Vaughan}(2013)}]{Vaughan:2013bca}%
  \BibitemOpen
  \bibfield  {author} {\bibinfo {author} {\bibfnamefont {S.}~\bibnamefont
  {Vaughan}},\ }\bibfield  {title} {\bibinfo {title} {{Random time series in
  Astronomy}},\ }\href {https://doi.org/10.1098/rsta.2011.0549} {\bibfield
  {journal} {\bibinfo  {journal} {Phil. Trans. Roy. Soc. Lond. A}\ }\textbf
  {\bibinfo {volume} {371}},\ \bibinfo {pages} {0549} (\bibinfo {year}
  {2013})},\ \Eprint {https://arxiv.org/abs/1309.6435} {arXiv:1309.6435
  [astro-ph.IM]} \BibitemShut {NoStop}%
\bibitem [{\citenamefont {Desherevskii}\ \emph {et~al.}(2017)\citenamefont
  {Desherevskii}, \citenamefont {Zhuravlev}, \citenamefont {Nikolsky},\ and\
  \citenamefont {Sidorin}}]{Desherevskii2017}%
  \BibitemOpen
  \bibfield  {author} {\bibinfo {author} {\bibfnamefont {A.~V.}\ \bibnamefont
  {Desherevskii}}, \bibinfo {author} {\bibfnamefont {V.~I.}\ \bibnamefont
  {Zhuravlev}}, \bibinfo {author} {\bibfnamefont {A.~N.}\ \bibnamefont
  {Nikolsky}},\ and\ \bibinfo {author} {\bibfnamefont {A.~Y.}\ \bibnamefont
  {Sidorin}},\ }\bibfield  {title} {\bibinfo {title} {Problems in analyzing
  time series with gaps and their solution with the winabd software package},\
  }\href {https://doi.org/10.1134/S0001433817070027} {\bibfield  {journal}
  {\bibinfo  {journal} {Izvestiya, Atmospheric and Oceanic Physics}\ }\textbf
  {\bibinfo {volume} {53}},\ \bibinfo {pages} {659} (\bibinfo {year}
  {2017})}\BibitemShut {NoStop}%
\bibitem [{\citenamefont {Lomb}(1976)}]{lomb1976least}%
  \BibitemOpen
  \bibfield  {author} {\bibinfo {author} {\bibfnamefont {N.~R.}\ \bibnamefont
  {Lomb}},\ }\bibfield  {title} {\bibinfo {title} {Least-squares frequency
  analysis of unequally spaced data},\ }\href@noop {} {\bibfield  {journal}
  {\bibinfo  {journal} {Astrophysics and space science}\ }\textbf {\bibinfo
  {volume} {39}},\ \bibinfo {pages} {447} (\bibinfo {year} {1976})}\BibitemShut
  {NoStop}%
\bibitem [{\citenamefont {Scargle}(1982)}]{scargle1982studies}%
  \BibitemOpen
  \bibfield  {author} {\bibinfo {author} {\bibfnamefont {J.~D.}\ \bibnamefont
  {Scargle}},\ }\bibfield  {title} {\bibinfo {title} {Studies in astronomical
  time series analysis. ii-statistical aspects of spectral analysis of unevenly
  spaced data},\ }\href@noop {} {\bibfield  {journal} {\bibinfo  {journal}
  {Astrophysical Journal, Part 1, vol. 263, Dec. 15, 1982, p. 835-853.}\
  }\textbf {\bibinfo {volume} {263}},\ \bibinfo {pages} {835} (\bibinfo {year}
  {1982})}\BibitemShut {NoStop}%
\bibitem [{\citenamefont {Nielsen}(2019)}]{nielsen2019practical}%
  \BibitemOpen
  \bibfield  {author} {\bibinfo {author} {\bibfnamefont {A.}~\bibnamefont
  {Nielsen}},\ }\href@noop {} {\emph {\bibinfo {title} {Practical time series
  analysis: Prediction with statistics and machine learning}}}\ (\bibinfo
  {publisher} {O'Reilly Media},\ \bibinfo {year} {2019})\BibitemShut {NoStop}%
\bibitem [{\citenamefont {Shumway}\ \emph {et~al.}(2000)\citenamefont
  {Shumway}, \citenamefont {Stoffer},\ and\ \citenamefont
  {Stoffer}}]{shumway2000time}%
  \BibitemOpen
  \bibfield  {author} {\bibinfo {author} {\bibfnamefont {R.~H.}\ \bibnamefont
  {Shumway}}, \bibinfo {author} {\bibfnamefont {D.~S.}\ \bibnamefont
  {Stoffer}},\ and\ \bibinfo {author} {\bibfnamefont {D.~S.}\ \bibnamefont
  {Stoffer}},\ }\href@noop {} {\emph {\bibinfo {title} {Time series analysis
  and its applications}}},\ Vol.~\bibinfo {volume} {3}\ (\bibinfo  {publisher}
  {Springer},\ \bibinfo {year} {2000})\BibitemShut {NoStop}%
\bibitem [{\citenamefont {VanderPlas}(2018)}]{VanderPlas_2018}%
  \BibitemOpen
  \bibfield  {author} {\bibinfo {author} {\bibfnamefont {J.~T.}\ \bibnamefont
  {VanderPlas}},\ }\bibfield  {title} {\bibinfo {title} {Understanding the
  {L}omb–{S}cargle periodogram},\ }\href
  {https://doi.org/10.3847/1538-4365/aab766} {\bibfield  {journal} {\bibinfo
  {journal} {The Astrophysical Journal Supplement Series}\ }\textbf {\bibinfo
  {volume} {236}},\ \bibinfo {pages} {16} (\bibinfo {year} {2018})}\BibitemShut
  {NoStop}%
\bibitem [{\citenamefont {{Astropy Collaboration}}\ \emph
  {et~al.}(2022)\citenamefont {{Astropy Collaboration}}, \citenamefont
  {{Price-Whelan}}, \citenamefont {{Lim}} \emph {et~al.}}]{astropy:2022}%
  \BibitemOpen
  \bibfield  {author} {\bibinfo {author} {\bibnamefont {{Astropy
  Collaboration}}}, \bibinfo {author} {\bibfnamefont {A.~M.}\ \bibnamefont
  {{Price-Whelan}}}, \bibinfo {author} {\bibfnamefont {P.~L.}\ \bibnamefont
  {{Lim}}}, \emph {et~al.},\ }\bibfield  {title} {\bibinfo {title} {{The
  Astropy Project}},\ }\href {https://doi.org/10.3847/1538-4357/ac7c74}
  {\bibfield  {journal} {\bibinfo  {journal} {apj}\ }\textbf {\bibinfo {volume}
  {935}},\ \bibinfo {eid} {167} (\bibinfo {year} {2022})},\ \Eprint
  {https://arxiv.org/abs/2206.14220} {arXiv:2206.14220 [astro-ph.IM]}
  \BibitemShut {NoStop}%
\end{thebibliography}%

\end{document}